\documentclass[2column]{aa} 
\usepackage{natbib}
\usepackage{graphicx}
\usepackage{txfonts}

\begin{document}

\title{Analytical Treatment of Planetary Resonances}  
\titlerunning{First Order Resonances}
\author{Konstantin Batygin\inst{1,2}\thanks{kbatygin@cfa.harvard.edu} \and Alessandro Morbidelli\inst{1}\thanks{morby@oca.eu}} 
\institute{Departement Lagrange, Observatoire de la C${\mathrm{\hat{o}}}$te d'Azur, 06304 Nice, France
\and
Institute for Theory and Computation, Harvard-Smithsonian Center for Astrophysics, 60 Garden St., Cambridge, MA, 02138 USA}
\abstract{An ever-growing observational aggregate of extrasolar planets has revealed that systems of planets that reside in or near mean-motion resonances are relatively common. While the origin of such systems is attributed to protoplanetary disk-driven migration, a qualitative description of the dynamical evolution of resonant planets remains largely elusive. Aided by the pioneering works of the last century, we formulate an approximate, integrable theory for first-order resonant motion. We utilize the developed theory to construct an intuitive, geometrical representation of resonances within the context of the unrestricted three-body problem. Moreover, we derive a simple analytical criterion for the appearance of secondary resonances between resonant and secular motion. Subsequently, we demonstrate the onset of rapid chaotic motion as a result of overlap among neighboring first-order mean-motion resonances, as well as the appearance of slow chaos as a result of secular modulation of the planetary orbits. Finally, we take advantage of the integrable theory to analytically show that, in the adiabatic regime, divergent encounters with first-order mean-motion resonances always lead to persistent apsidal anti-alignment.}
\maketitle

\section{Introduction}

The continued search for extrasolar planets around nearby stars has proven to be a goldmine of discoveries in numerous sub-fields of planetary astrophysics. Among the disciplines that have benefited the most is the study of orbital dynamics, as the aggregate of known planetary system architectures has grown immensely. Importantly, the observations collectively suggest that the orbital structure of the solar system is a singular example among numerous possible dynamical states. Indeed, orbital configurations that are quite unlike our own exist. Within the currently available observational collection, of particular interest is the class of systems that contains planets that reside in or near mean-motion resonances or loosely speaking, display integer commensurabilities among the orbital periods. 

The range of parameter space occupied by resonant planets is remarkably vast. Long-term radial velocity monitoring has revealed that giant planets occasionally reside in mean motion resonances at orbital distances exceeding $\sim 1$AU \citep{2011ApJ...730...93W}. At the same time, searches aimed at transiting exoplanets (the \textit{Kepler} mission in particular) have shown that (near-)resonances are quite common among low-mass planets that reside in close proximity to their host stars \citep{2012arXiv1202.6328F}. Furthermore, it has been proposed that the giant planets of the solar system once occupied a resonant state 
\citep{2001MNRAS.320L..55M,2007AJ....134.1790M}, before undergoing a transient dynamical instability that drove the orbits to their current locations \citep{2010ApJ...716.1323B,2011AJ....142..152L}.

The prevalence of mean motion commensurabilities among planets is probably not coincidental and is likely to be a result of a physical mechanism. Indeed, it is believed that resonances congregate at an epoch in the dynamical evolution when the protoplanetary nebula is still present. Specifically, interactions between newly formed planets and the gaseous disk, into which they are embedded, leads to a time-irreversible exchange of angular momentum that results in planetary migration \citep{1980ApJ...241..425G,1996Natur.380..606L,2007A&A...461.1173C}. Although the particular regime (i.e. rate, direction) of the migration depends upon the planetary mass \citep{2010apf..book.....A} as well as the thermodynamic properties of the disk \citep{2009MNRAS.394.2283P,2011A&A...536A..77B}, occurrences where migration among planetary pairs is slow and convergent are thought to be common \citep{2007ApJ...654.1110T}. In such cases, provided that the disk in question is not overwhelmingly turbulent \citep{2008ApJ...683.1117A,2008A&A...482..677C,2011ApJ...726...53K} and the planetary orbits are nearly circular, capture into resonance is essentially guaranteed \citep{1982CeMec..27....3H, 1986sate.conf..159P}. 

\begin{figure}
\includegraphics[width=1\columnwidth]{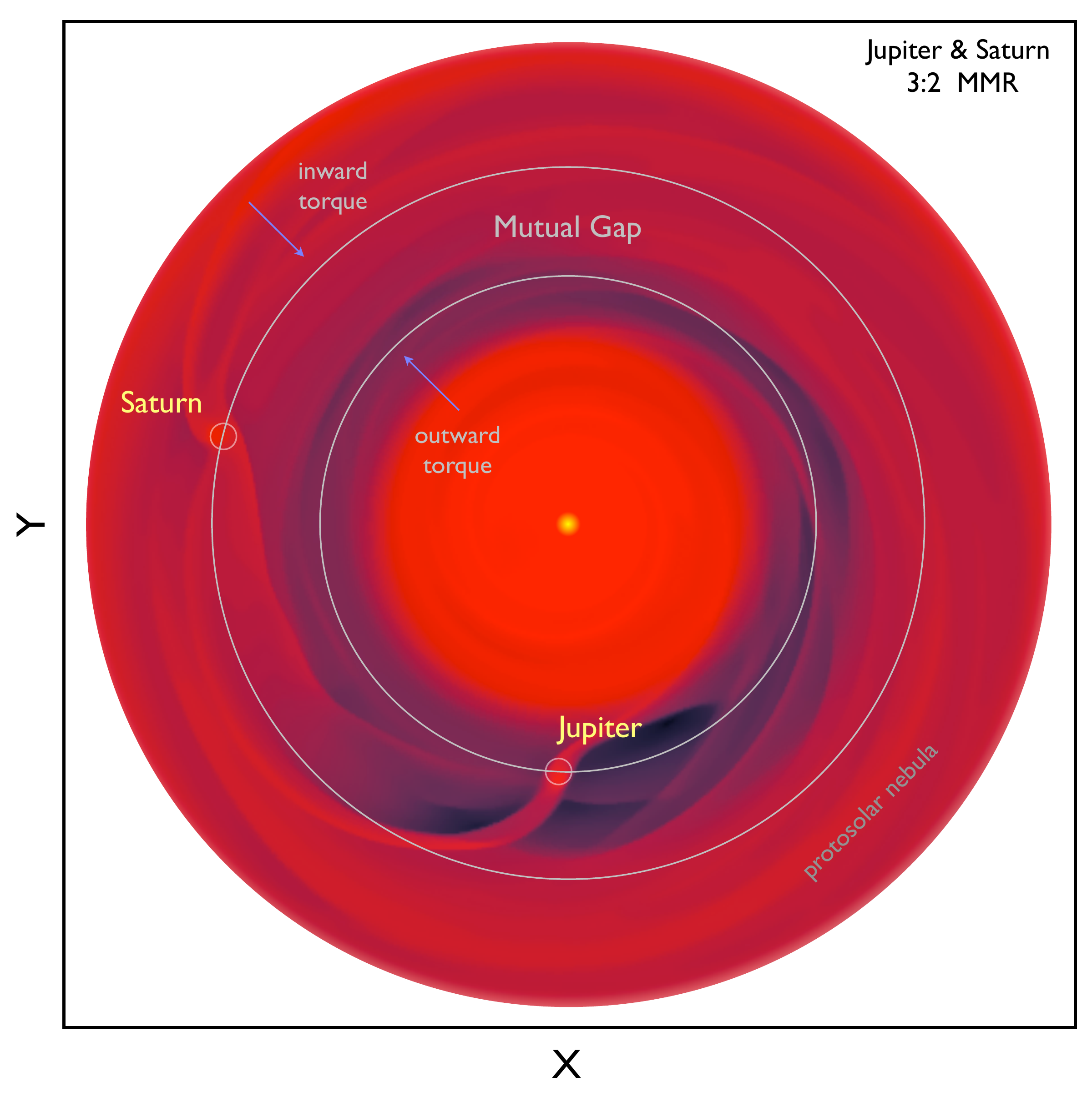}
\caption{Jupiter and Saturn in a resonant configuration. The figure shows a snapshot of the orbital state of Jupiter and Saturn, embedded into the solar nebula. The planets are locked in a 3:2 mean motion resonance. By virtue of carving out a mutual gap in the gaseous disk, the planetary migration direction is reversed to point outwards. The locations of the planets as well as their orbits are labeled accordingly and the background color represents the logarithm of the gas density. This simulation is a reproduction of the numerical experiments performed by \citet{2001MNRAS.320L..55M, 2007Icar..191..158M}. The snapshot shows the system at a time when Jupiter is at 4.3 AU.}
\label{jsgas}
\end{figure}

An example of resonant capture among giant planets, resulting from disk-dirven migration is shown in Figure (\ref{jsgas}). Specifically, the figure shows Jupiter and Saturn locked in a 3:2 mean motion resonance, having opened a mutual gap in the protoplanetary disk. The figure depicts a reproduction of the results of \citet{2001MNRAS.320L..55M} and \citet{2007Icar..191..158M}, where all simulation parameters were adopted from the latter study. 

It is noteworthy that gaseous protoplanetary disks are not the only environments where migrating planets can encounter mean motion resonances. Massive objects embedded in debris disks often undergo planetesimal-driven migration \citep{1984Icar...58..109F,1998Sci...279...69M,2009Icar..199..197K}. In fact, \citet{1995AJ....110..420M} proposed exactly this process for the origin of the 3:2 mean motion resonance between Neptune and Pluto. 

Yet another setting where resonant encounters are common is the orbital region occupied by planetary satellites \citep{1986sate.conf..159P,1999ARA&A..37..533P}. In the context of the planetary satellite problem, migration is usually forced by tidal interactions with the host planet \citep{1963MNRAS.126..257G, 1966Icar....5..375G}. An oft-quoted example of a tidally assembled system is the Laplace resonance of the Galilean moons \citep{1966Icar....5..375G, 1983Icar...53...55H}. Systems of resonant planets on orbits that are close to their host stars also interact with the star tidally. However, in such systems, the interplay between the resonant dynamics and the dissipative forces results in a repulsion of the orbits \citep{2012arXiv1204.2791B, 2012ApJ...756L..11L}, rather than a convergence towards nominal commensurability.

Quite contrary to the examples described above, encounters with mean motion resonances by divergently migrating planets can never result in capture \citep{1991pscn.proc..193H,1999ssd..book.....M}. Instead, passage through resonance leads to an impulsive excitation of the orbital parameters. As an example, such a process is thought to be responsible for the mutual inclinations of the Uranian satellites \citep{1988Icar...74..153P,1990Icar...85..394T}. Furthermore, our own Jupiter and Saturn may have once encountered the 2:1 mean motion resonance, jumpstarting the transient dynamical instability of the solar system that helped shape the Kuiper belt \citep{2005Natur.435..459T, 2008Icar..196..258L, 2011ApJ...738...13B}. 

The long-term evolution of resonant objects can be quite complex. In fact, it is now well known that overlap of resonances gives rise to chaos \citep{1979PhR....52..263C, 1980AJ.....85.1122W}. In turn, this can result in orbital instabilities. Indeed, the process of chaotic clearing of resonant orbits is illustrated by the lack of objects in the Kirkwood gaps of the Asteroid belt \citep{1983Icar...56...51W, 1990CeMDA..47...99H, 1997AJ....114.1246M}.

\begin{figure}
\includegraphics[width=1\columnwidth]{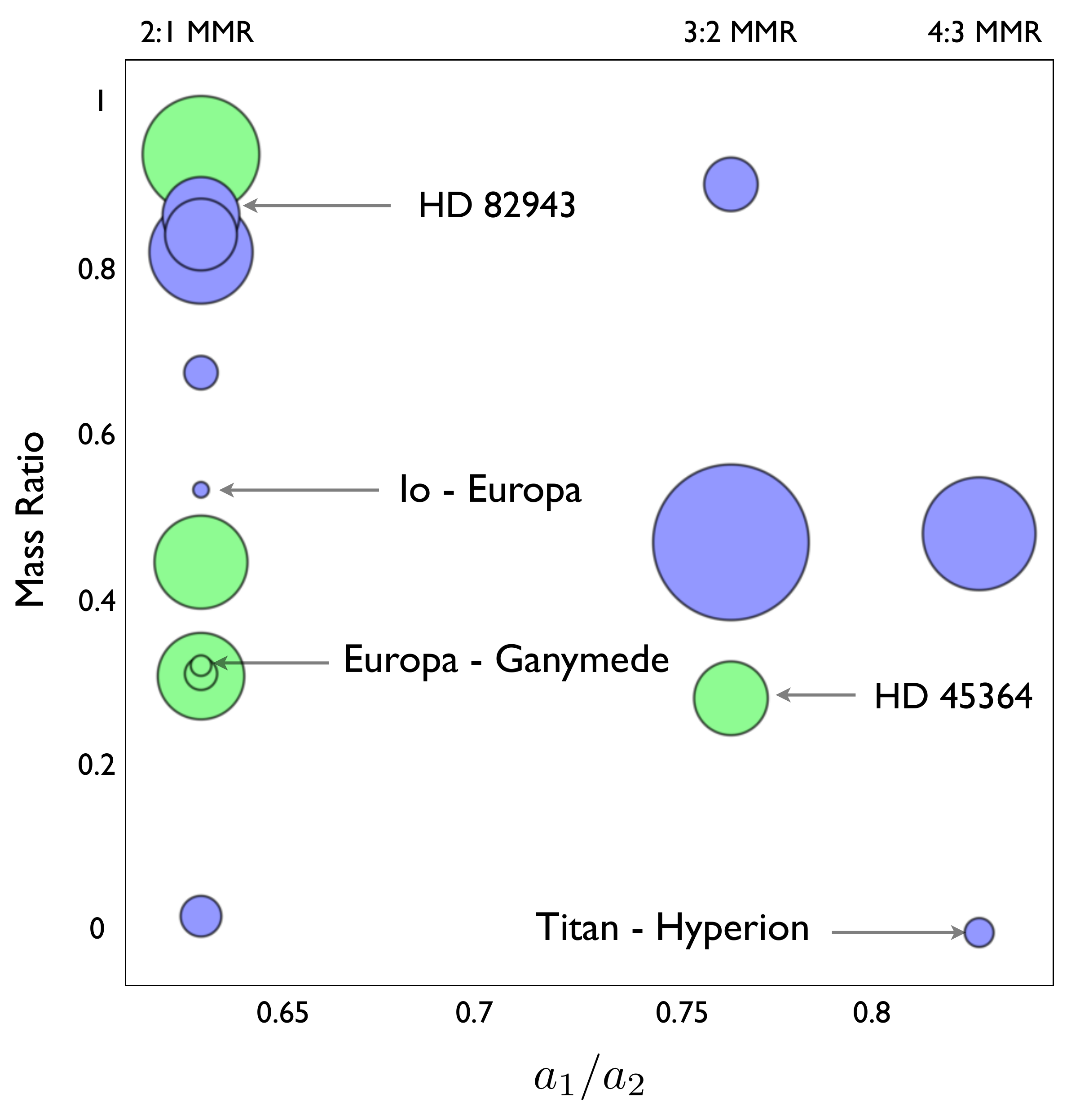}
\caption{Mass ratio of resonant exoplanets and some solar system satellite pairs. Only well-characterized systems \citep{2011ApJ...730...93W} are listed. The size of each bubble is indicative of the orbital distance of the inner orbit in units of the central body radii. Systems depicted by blue circles are those with a more massive inner object. Systems represented with green circles are those with a more massive outer object. This figure clearly shows that the restricted formalism of the three-body problem is inapplicable in numerous settings of astrophysical interst.}
\label{mratio}
\end{figure}

The majority of the work on the chaotic dynamics of mean motion resonances has found its application in the study of the orbital evolution of small bodies with negligible masses (e.g. Asteroids, Kuiper Belt objects, (ir)regular satellites) \citep{2002aste.conf..379N, 2008ssbn.book..275M}, although chaotic diffusion of planetary orbits in the outer solar system has also received some attention \citep{1999Sci...283.1877M}. With a growing aggregate of detected extrasolar planets, (near-)resonant planetary pairs characterized by secondary mass-ratios close to unity have become common. This implies an expanded tally of objects to which the well-studied restricted formalism, where one of the three bodies is taken to be mass-less, is inapplicable. In particular, Figure (\ref{mratio}) depicts the mass ratios of the currently known, well-characterized first-order resonant extrasolar planets \citep{2011ApJ...730...93W} as well as some solar system examples. The sizes of the circles are representative of the planetary orbital radii in units of the primary's physical radius. Green circles denote resonant pairs with a more massive outer planet while blue circles denote systems with a more massive inner planet.

Influenced by the emergence of observational detections, a handful of authors have studied the global resonant dynamics of the unrestricted three-body problem \citep{2005ApJ...634..625R,2007CeMDA..98....5C, 2008MNRAS.387..747M}. While quantitatively precise, the latter studies are generally tailored to particular systems, characterized by specific resonances and mass ratios. This renders the translation of the results to other systems and the acquisition of an overall understanding of the motion a difficult task. Indeed, a more physically intuitive and broadly applicable picture of resonant dynamics is desirable. 

Here, we shall set out to draw such a picture. As such, the analytical characterization of the global first-order resonant dynamics, is the primary purpose of this work.The paper is organized as follows. In the following section, we formulate a fully analytical, integrable treatment of resonant phenomena. Using the approximate theory, we construct surfaces of section that prove useful as a visual representation of the resonant motion. In section 3, we consider the onset of chaos via overlap of neighboring resonances as well as the incorporation of higher-order secular perturbations into the developed framework. In section 4, we apply the constructed formalism to divergent resonant encounters and examine the characteristic features of post-encounter dynamical states. We summarize and discuss our results in section 5.

\section{Resonant Motion}

\subsection{An Integrable Approximation}

Our first aim is to construct an integrable approximation to the first-order resonant motion (i.e. the orbital period ratio $P_1/P_2 \approx k/(k-1), k \in \mathbb{Z}$) of two massive secondary bodies with masses $m_1$ and $m_2$, which orbit a much more massive ($M \gg m_{1,2}$) primary. By convention, we shall take the subscripts $1$ and $2$ to denote the inner and outer orbits respectively. The exact Hamiltonian, $\mathcal{H}$ which governs the gravitational three body problem is characterized by six degrees of freedom. Specifically, the canonical heliocentric formulation of $\mathcal{H}$ reads \citep{1902BuAsI..19..289P, 1995CeMDA..62..193L, 2002mcma.book.....M}:
\begin{eqnarray}
\label{Hbasic}
\mathcal{H} = \frac{M + m_1}{2 M }\frac{p_1^2}{ m_1} + \frac{M + m_2}{2 M }\frac{p_2^2}{m_2} - \frac{\mathcal{G} M m_1}{r_1}  \nonumber \\
 -  \frac{\mathcal{G} M m_2}{r_2} + \frac{p_1 \cdot p_2}{M} - \mathcal{G} \frac{m_1 m_2}{\Delta_{12}}
\end{eqnarray}
where $\mathcal{G}$ is the gravitational constant, $p$ is the barycentric linear momentum, $r_1$, $r_2$ are the distances between the primary and the secondaries while $\Delta_{12}$ is the distance between the planets. Today, the availability of numerical tools for integration of the Hamiltonian (\ref{Hbasic}) \citep{1991AJ....102.1528W,1998AJ....116.2067D,1999MNRAS.304..793C} allows for a prompt and precise realization of a given system's orbital evolution. However, any such realization provides a scarce theoretical basis for the characterization of the dynamics. Moreover, as was first pointed out by \citet{1902BuAsI..19..289P}, such solutions may exhibit chaotic motion further obscuring candid interpretation. Consequently, rather than working with the Hamiltonian (\ref{Hbasic}) directly, it is sensible to turn to the classical perturbation methods developed over the last four centuries, in search of a suitable approximation to the Hamiltonian (\ref{Hbasic}). 

Throughout the following derivation, we shall be aided by numerous preceding contributions to the study of resonance in celestial mechanics. Specifically, we shall follow the pioneering work of \citet{1976ARA&A..14..215P} and \citet{1984CeMec..32..307S}. The calculation will be greatly simplified by a reducing transformation (see \citet{1986CeMec..38..335H, 1986CeMec..38..175W}) and the final Hamiltonian will closely resemble the second fundamental model for resonance \citep{1983CeMec..30..197H}.

It is useful to begin, (without fear of overstating the obvious) by pointing out that the combination of the first and third as well as second and fourth terms in equation (\ref{Hbasic}) govern the Keplerian motion of the planets. It can be easily shown \citep{1999ssd..book.....M,2002mcma.book.....M} that in terms of orbital elements, this Keplerian part of the Hamiltonian can be written as follows:
\begin{equation}
\label{Hkeporbel}
\mathcal{H}_{\rm{kep}} = - \frac{\mathcal{G} M m_1}{2 a_1} - \frac{\mathcal{G} M m_2}{2 a_2},
\end{equation}
where $a$ is the semi-major axis. The remaining terms in the Hamiltonian (\ref{Hbasic}) govern the planet-planet interactions and are much smaller in magnitude. Accordingly, it is often called the disturbing function, since it provides small perturbations to the integrable Hamiltonian (\ref{Hkeporbel}) that are still important in the long term.

A qualitative analysis of the dynamics can be performed by expanding the disturbing function as a Fourier series in the orbital angles and a power series of the planetary eccentricities and inclinations \citep{1995CeMDA..62..193L, 2010A&A...522A..60L}. Accordingly, this procedure allows for the identification of resonant terms, that is, harmonics that vary on a timescale much longer than the orbital timescale in the vicinity of exact commensurability. While such terms are dynamically important and should be retained in the Hamiltonian, short-periodic terms (i.e. those that vary on an orbital timescale) can be readily averaged over and dropped from the Hamiltonian \citep{1999ssd..book.....M}. 

It is noteworthy that in addition to short-periodic and resonant terms, the disturbing function also contains secular terms which do not depend on the mean longitudes of the planets. The leading secular terms are of order $\mathcal{O} (e^2 , i^{2})$, where $e$ and $i$ are the eccentricity and inclination respectively. For the purposes of the construction of a first-order resonant theory, we shall neglect them, along with all resonant terms of order greater than unity in $e$ and $i$. Additionally, we shall only retain terms that are linear in planetary masses. However, as will be shown in the subsequent sections, these higher-order terms play a crucial role in the onset of chaotic motion. 

In accord with the above-mentioned linear expansion of the disturbing function, we can approximate the Hamiltonian (\ref{Hbasic}) as 
\begin{eqnarray}
\mathcal{H} \simeq \mathcal{H}_{\rm{kep}} + \mathcal{H}_{\rm{res}} + \mathcal{O} (e^2 , i^{2}),
\end{eqnarray}
where the $k:k-1$ resonant perturbation to the Keplerian motion reads:
\begin{eqnarray}
\label{Hkepres}
\mathcal{H}_{\rm{res}} = - \frac{\mathcal{G} m_1 m_2}{a_2} ( f_{\rm{res}}^{(1)} e_1 \cos(k \lambda_2 - (k-1) \lambda_1 - \varpi_1)  \nonumber \\
+ f_{\rm{res}}^{(2)} e_2 \cos(k \lambda_2 - (k-1) \lambda_1 - \varpi_2) ).
\end{eqnarray}
Following conventional notation, $\varpi$ denotes the longitude of perihelion and $\lambda = \mathcal{M}+\varpi$ is the mean longitude, $\mathcal{M}$ being the mean anomaly. The quantities $f_{\rm{res}}^{(1)}$ and $f_{\rm{res}}^{(2)}$ are of order unity to within a factor of a few and (weakly) depend on the semi-major axis ratio $(a_1/a_2)$ only. Their values are tabulated in numerous references and can be easily evaluated numerically with the aid of computer algebra (see for example \citet{2007CeMDA..98....5C, 2010A&A...522A..60L}). 

At the expense of working in a noninertial reference frame, we had to introduce the indirect term, $p_1 \cdot p_2/M$, into the disturbing function that accounts for fixing the origin on the central body \citep{2010A&A...522A..60L}. However, this correction is trivial, given that all indirect terms in $\mathcal{H}_{\rm{res}}$ corresponding to the same harmonics have the same dependence on the actions as the direct ones, meaning that the indirect terms can be accounted for simply by modifying the coefficients $f_{\rm{res}}$.

\begin{figure*}
\includegraphics[width=1\textwidth]{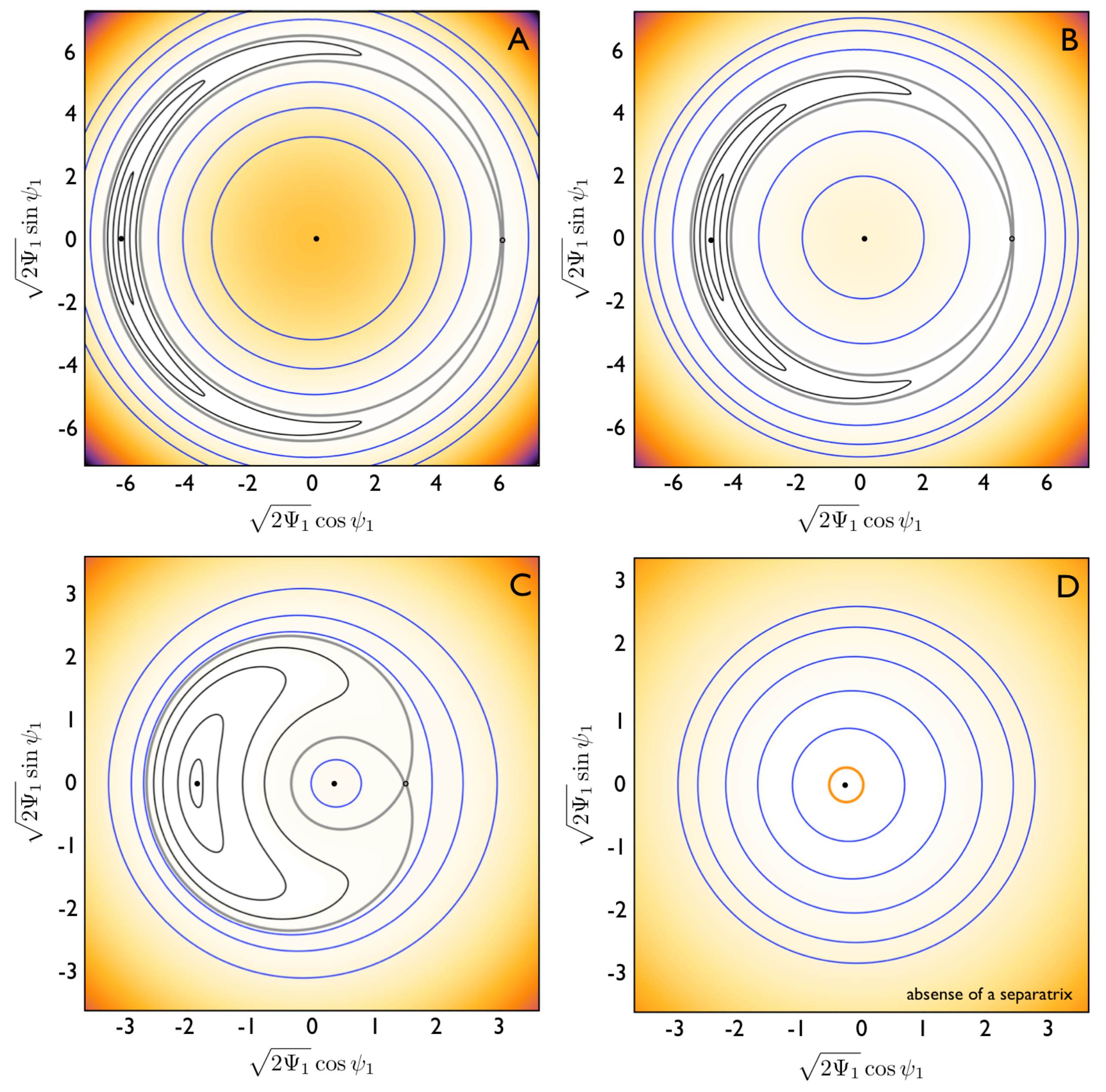}
\caption{Level curves of the Hamiltonian (\ref{Honedof}). The plotted energy levels correspond to the Hamiltonian characterized by the mass-ratio and angular momentum equivalent to that of the 2:1 resonant HD 82943 system (see Table 1 as well as Figure (\ref{Poincare21})). The four dynamical portraits depicted in panels labeled A B C D exhibit different proximities to exact resonance. Specifically, the associated values of $\Phi_2$ are A : $\Phi_2 = 1.6 \times 10^{-4}$, B : $\Phi_2 = 4.8 \times 10^{-4}$, C : $\Phi_2 = 9.6 \times 10^{-4}$, D : $\Phi_2 = 1.12 \times 10^{-3}$. Note that the dynamical portraits shown in panels A, B and C feature the presence of separatrcies, shown as thick gray curves. The separatrix disappears in panel D, although libration of the critical angle is still possible, as depicted by the thick orange circle. Note the factor of two reduction in the axes ranges between panels A \& B and C \& D.}
\label{levelcurves}
\end{figure*}

Although the functional form of the simplified Hamiltonian is given, Keplerian orbital elements do not form a canonically conjugated set. Thus to make further progress, we convert to Poincar$\acute{\rm{e}}$ action-angle variables defined as:
\begin{eqnarray}
\Lambda = \mu \sqrt{\mathcal{G} (M+m) a}, \ \ \ \ \ \ \ \ \ \ \ \ \ \ \ \ \ \ \ \ \ \ \lambda = \mathcal{M} + \varpi, \\
\Gamma = \Lambda (1 - \sqrt{1-e^2}) \approx \Lambda \ e^2/2, \ \ \ \ \ \ \ \ \ \  \ \ \gamma = - \varpi,
\end{eqnarray}
where $\mu = m M / (M+m) \simeq m$ is the reduced mass. In terms of the Poincar$\acute{\rm{e}}$ variables, the Hamiltonians, $\mathcal{H}_{\rm{kep}}$ and $\mathcal{H}_{\rm{res}}$ take on the following forms respectively \citep{1999ssd..book.....M}:
\begin{equation}
\label{Hkeplambda}
\mathcal{H}_{\rm{kep}} = - \frac{\mathcal{G}^2 M^2 m_1^3}{2 \Lambda_1^2} - \frac{\mathcal{G}^2 M^2 m_2^3}{2 \Lambda_2^2},
\end{equation}
\begin{eqnarray}
\label{Hrespoincare}
\mathcal{H}_{\rm{res}} = - \frac{\mathcal{G}^2 M m_1 m_2^3}{\Lambda_2^2} ( f_{\rm{res}}^{(1)} \sqrt{\frac{2 \Gamma_1}{\Lambda_{1}}} \cos(k \lambda_2 - (k-1) \lambda_1 + \gamma_1)  \nonumber \\
+ f_{\rm{res}}^{(2)}  \sqrt{\frac{2 \Gamma_2}{\Lambda_{2}}} \cos(k \lambda_2 - (k-1) \lambda_1 + \gamma_2) ).
\end{eqnarray}
Although we did not explicitly assume coplanar orbits, the linear expansion of the disturbing function contains no terms that depend on the longitudes of the ascending node (the third Poincar$\acute{\rm{e}}$ angle), or the orbital inclinations (related to the third Poincar$\acute{\rm{e}}$ action). This renders these quantities integrals of motion. Therefore, it is evident that the number of degrees of freedom of $\mathcal{H}$ has been reduced to four (although the distinct presence of only two harmonics in equation (\ref{Hrespoincare}) suggests that $\mathcal{H}$ can be reduced to a two degrees of freedom system with ease).

Because we are interested in near-commensurate planetary motion, it is sensible to expand the Keplerian Hamiltonian around the nominal resonant location. Carrying out the expansion to second order in $\delta \Lambda = \Lambda - [\Lambda]$, where $[\Lambda]$ is the nominal resonant value of $\Lambda$, we have:
\begin{equation}
\label{nominalexpand}
-\frac{\mathcal{G}^2 M^2 m^3}{2 \Lambda^2} \simeq - \frac{G^2 M^2 m^3 }{2 [\Lambda]^2} + \frac{G^2 M^2 m^3}{[\Lambda]^3}\delta \Lambda - \frac{3 G^2 M^2 m^3}{2 [\Lambda]^4}\delta \Lambda^2.
\end{equation}

Substituting the definition of $\delta \Lambda$ into equation (\ref{nominalexpand}) and dropping the dynamically unimportant constant terms, $\mathcal{H}_{\rm{kep}}$ takes on the following remarkably simple form:
\begin{equation}
\label{HkepLambd}
\mathcal{H}_{\rm{kep}} = 4([n]_1 \Lambda_1 + [n]_2 \Lambda_2) - \frac{3}{2}([h]_1 \Lambda_1^2 +  [h]_2 \Lambda_2^2).
\end{equation}
Here, $[n] = \sqrt{\mathcal{G} M/[a]^3}$ is the nominal mean motion and $[h] = [n]/[\Lambda]=1/(m[a]^2)$. In accord with the above approximation, we shall also evaluate $\mathcal{H}_{\rm{res}}$ at $[\Lambda]$, as it is already of order $\mathcal{O}(e)$. Indeed, this step is relevant to the evaluation of the resonant strengths, as the coefficients $f_{\rm{res}}$ can now be considered truly constant. 

In the limit of very small eccentricities and semi-major axes ratios exceeding the nominal resonant value (that is, assuming that the system remains close to the (pseudo-)resonant equilibrium point, which is in turn taken to be close to the origin of the phase-space portrait), the dynamics governed by Hamiltonians (\ref{HkepLambd}) and (\ref{Hrespoincare}) can be treated linearly. An analysis of this kind has recently been performed by \citet{2012arXiv1204.2791B} and the resulting equations were used to study the resonant evolution of close-in planets under the effect of tides. In this work, we wish to provide a more general picture of resonant motion that is not limited to the vicinity of any equilibrium point. Consequently, here, we retain the nonlinear coupling of the actions inherent to (\ref{Hrespoincare}).

The functional form of the resonant harmonics can be simplified considerably by employing a canonical contact transformation of variables, arising from the following generating function of the second kind \citep{1984CeMec..32..307S}:
\begin{equation}
F_2 = \lambda_1 \mathcal{K} + (k \lambda_2 - (k-1) \lambda_1)\Theta. % + \gamma_1\Gamma_1 +  \gamma_2\Gamma_2
\end{equation}
An application of the transformation equations $\Lambda =\partial F_2/\partial \lambda$ yields new action-angle variables:
\begin{eqnarray}
\label{transformone}
\mathcal{K} = \Lambda_1 + \frac{k -1}{k} \Lambda_2, \ \ \  \ \ \ \ \ \ \ \ \ \ \ \ \ \ \ \  \ \ \  \ \kappa = \lambda_1 \nonumber, \\
\Theta = \Lambda_2 / k,   \ \ \ \ \ \ \ \ \ \ \ \ \ \ \ \ \ \   \theta = k \lambda_2 - (k-1) \lambda_1.
\end{eqnarray}

Upon substitution of the new variables into the Hamiltonian and utilizing the resonant relationship $(k-1) [n]_1 = k [n]_2$, the Keplerian Hamiltonian becomes
\begin{eqnarray}
\label{Hkeptheta}
\mathcal{H}_{\rm{kep}} = 4 [n]_1 \mathcal{K}  + 3[h]_1 (k-1) \mathcal{K} \Theta \nonumber \\
-\frac{3}{2}\left([h]_1(k-1)^2 + [h]_2 k^2  \right) \Theta^2 - \frac{3}{2} [h]_1 \mathcal{K}^2
\end{eqnarray}
Meanwhile, the resonant contribution to $\mathcal{H}$ now takes the form:
\begin{equation}
\label{Hrestheta}
\mathcal{H}_{\rm{res}} =- \alpha \sqrt{2 \Gamma_1} \cos(\gamma_1 + \theta) - \beta \sqrt{2 \Gamma_2} \cos(\gamma_2 + \theta),
\end{equation}
where
\begin{eqnarray}
\label{alpha}
\alpha =  \frac{\mathcal{G}^2 M m_1 m_2^3}{[\Lambda]_2^2} \frac{ f_{\rm{res}}^{(1)}}{\sqrt{[\Lambda]_{1}}}, \nonumber \\ \nonumber \\
\beta = \frac{\mathcal{G}^2 M m_1 m_2^3}{[\Lambda]_2^2} \frac{ f_{\rm{res}}^{(2)}}{\sqrt{[\Lambda]_{2}}}.
\end{eqnarray}
Note that $\kappa$ is no longer present in $\mathcal{H}$. Thus, $\mathcal{K}$ is now a constant of motion and the number of degrees of freedom of $\mathcal{H}$ has been reduced to three. Physically, the conservation of $\mathcal{K}$ arises from the fact that the resonant Hamiltonian depends on the semi-major axis ratio rather than the semi-major axes themselves (recall that $\mathcal{K} \propto \Lambda \propto \sqrt{a}$). Accordingly, \citet{2008MNRAS.387..747M} have dubbed $\mathcal{K}$ a ``scaling parameter" (see also their discussion of $\mathcal{K}$'s significance and its relationship to the behavior of the semi-major axes outside the resonant domain).

The utility of $\mathcal{K}$ can be illustrated intuitively by expressing it in a dimensionless form:
\begin{equation}
\label{Kdimless}
\frac{\mathcal{K}}{\Lambda_2} = \frac{m_1}{m_2} \sqrt{\frac{a_1}{a_2}} + \frac{k-1}{k}.
\end{equation}
For a given mass ratio, we can choose a nominal semi-major axis ratio, $[a]_1/[a]_2$ and obtain the value of $\mathcal{K}/[\Lambda]_2$ accordingly. Although $\mathcal{K}$ is simply a constant of motion and can in principle take on arbitrary values, without loss of generality, we can choose $[\Lambda]_2 = 1$, thereby defining a natural value of $\mathcal{K}$. In this sense, the actual value of $\mathcal{K}$ is simply representative of the units in which the semi-major axes are measured. Once the value of $\mathcal{K}$ is fixed, both planetary semi-major axes, $a_1$ and $a_2$ are unequivocally defined given their ratio, $a_1/a_2$.

The conservation of $\mathcal{K}$ is of additional importance, as it yields the location of nominal mean motion, $[n]$, around which we have chosen to expand the Keplerian Hamiltonian (\ref{Hkeplambda}). In the unrestricted problem, the semi-major axes of both planets deviate away from nominal commensurability during a resonant cycle. The extent of such deviation is dependent on the planetary masses. Thus, it would seem that the nominal locations of the semi-major axes are not defined a-priori. This issue is remedied by the fact that $\mathcal{K}$ encapsulates the planetary mass ratio. Consequently, given an (observed) pair of semi-major axes, the conserved value of $\mathcal{K}$ can be used to compute their nominal counterparts by setting $a_2 = ((k-1)/k)^{2/3} a_1$ in equation (\ref{Kdimless}).

\begin{table*}
\begin{center}
\begin{tabular}{c c c c c c c c c c c c c c c c} % left columns (4 columns)\\ 
$\rm{System}$ &  $M_{\star}$ ($M_{\odot}$)  & $m_1$ ($M_{\rm{Jup}}$)  &  $m_2$ ($M_{\rm{Jup}}$)  &  $a_1$ (\rm{AU})   &  $a_1$ (\rm{AU})  &  $e_1$  &  $e_2$  &  $\varpi_1$ (\rm{deg})  &  $\varpi_2$ (\rm{deg})  &  $\lambda_1$ (\rm{deg})  &  $\lambda_2$ (\rm{deg})  &  \\
\hline
$\rm{HD 82943}$ & $1.15$  & $2.01$ & $1.75$  & $0.752$ & $1.190$ & 0.42 & 0.14 & 122 & 130 & 119 & 122  \\
$\rm{HD 45364}$ &   0.82    & $0.19$  & $0.66$ & $0.681$  & 0.897 & 0.17 & 0.1 & 162 & 7 & 106 & 270   \\ 
\hline
\end{tabular}
\end{center}
\caption{Adopted orbital parameters of the systems considered in this work. The fits to radial velocity data for HD 82943 (2:1 resonance) and HD 45364 (3:2 resonance) were taken from Fit II of \citep{2006ApJ...641.1178L}, although the planetary masses correspond to Fit I, and \citep{2009A&A...496..521C} respectively. It is implicitly assumed that the system is coplanar and the minimum masses derived from the data are representative of the real planetary masses. Although it is understood that different but equally permissible fits to the data can be found, here we retain the quoted values, as we only wish to use the systems as illustrative examples. Moreover, we note that (at least in the case of HD 82943) the quoted elements are represented in Jacobi coordinates, although in practice, their interpretation as Poincar$\acute{\rm{e}}$ coordinates only introduces negligible errors. }
\end{table*}

For further calculations, we shall drop the first and last terms in the Hamiltonian (\ref{Hkeptheta}) because they are constant. Doing so does not change the overall picture of the dynamics with the exception of the eliminated ability to derive the time evolution of the individual mean longitudes, $\lambda_1$ and $\lambda_2$. However, all other information, including the behavior of the resonant critical arguments in equation (\ref{Hrespoincare}), is retained despite this simplification.

The Hamiltonians (\ref{Hkeptheta}) and (\ref{Hrestheta}) are equivalent to (in fact, a trivial transformation away from) those considered by \citet{1984CeMec..32..307S}. Through a rather involved calculation, utilizing the perturbation method devised by \citet{1966PASJ...18..287H}, these authors demonstrated the integrability of the first-order resonant motion. In a subsequent effort, aimed primarily at the elliptic restricted three-body problem, \citet{1986CeMec..38..335H} and \citet{1986CeMec..38..175W} greatly simplified the \citet{1984CeMec..32..307S} solution by introducing a canonical transformation that explicitly identifies a novel constant of motion. Here, we shall follow the latter approach.

Turning our attention to the resonant contribution to $\mathcal{H}$, let us transform the remaining Poincar$\acute{\rm{e}}$ variables ($\Gamma, \gamma$) to mixed secular cartesian coordinates \citep{1999ssd..book.....M}
\begin{eqnarray}
\label{secularcart}
x_1 = \sqrt{2 \Gamma_1} \cos(\gamma_1), \ \ \ \ \ y_1 = \sqrt{2 \Gamma_1} \sin(\gamma_1), \nonumber \\
x_2 = \sqrt{2 \Gamma_2} \cos(\gamma_2), \ \ \ \ \ y_2 = \sqrt{2 \Gamma_2} \sin(\gamma_2),
\end{eqnarray}
where $y$ is identified as the coordinate and $x$ is the conjugate momentum. After some manipulation, the resonant Hamiltonian reads:
\begin{equation}
\label{Hresseccart}
\mathcal{H}_{\rm{res}} = - (\alpha x_1 + \beta x_2) \cos(\theta) + (\alpha y_1 + \beta y_2) \sin(\theta).
\end{equation}

At this point, the Hamiltonian (\ref{Hresseccart}) is ready for yet another change of variables. In particular, we introduce the rotation formulated by \citet{1986CeMec..38..335H, 1986CeMec..38..175W} 
\begin{eqnarray}
\label{suicide}
u_1 = \frac{\alpha x_1 + \beta x_2}{\sqrt{\alpha^2+\beta^2}}, \ \ \ \ \ v_1 = \frac{\alpha y_1 + \beta y_2}{\sqrt{\alpha^2+\beta^2}}, \nonumber \\
u_2 = \frac{\beta x_1 - \alpha x_2}{\sqrt{\alpha^2+\beta^2}}, \ \ \ \ \ v_2 = \frac{\beta y_1 - \alpha y_2}{\sqrt{\alpha^2+\beta^2}}.
\end{eqnarray}
The canonical nature of this transformation can be verified by the Poisson bracket criterion $\{v_i,u_j\}_{(y_j,x_j)} = \delta_{i,j}$ \citep{2002mcma.book.....M}. Upon doing so, we can immediately identify $v$ as the coordinate and $u$ as the conjugated momentum.

Defining implicit action-angle polar coordinates ($\Phi, \phi$) as
\begin{equation}
\label{suicidepolar}
u = \sqrt{2 \Phi} \cos (\phi), \ \ \ \ \ v = \sqrt{2 \Phi} \sin (\phi),
\end{equation}
we can re-write the expression for the Hamiltonian (\ref{Hresseccart}) in a substantially simplified form:
\begin{equation}
\label{Hressuicidepolar}
\mathcal{H}_{\rm{res}} = - \sqrt{\alpha^2+\beta^2} \sqrt{2 \Phi_1} \cos(\phi_1 + \theta).
\end{equation}

Note that the Hamiltonian (\ref{Hressuicidepolar}) does not depend on $\phi_2$. Additionally, recall that effectively, the Keplerian Hamiltonian (\ref{Hkeptheta}) only depends on $\Theta$. Evidently, the newly defined action $\Phi_2$ is another constant of motion and the number of degrees of freedom of $\mathcal{H}$ has been reduced to two. Reverting to the original variables, the conservation of $\Phi_2$ implies that
\begin{equation}
e_1^2 \beta'^2 + e_2^2 \alpha'^2 -2 e_1 e_2 \alpha' \beta' \cos(\varpi_1 - \varpi_2) = \rm{const.}
\end{equation}
where $\alpha' = \alpha \sqrt{[\Lambda]_2}$ and $\beta' = \beta \sqrt{[\Lambda]_1} $.

We are only one degree of freedom away from integrability. Fortunately, the final reduction is arguably the simplest. However, before proceeding to the final transformation, let us briefly digress and rescale the total Hamiltonian. An examination of the expressions (\ref{Hkeptheta}) and (\ref{Hressuicidepolar}) reveals that currently, $\mathcal{H}$ is parameterized by $\alpha$, $\beta$, $[h]_1$ and $[h]_2$. Following \citet{1983CeMec..30..197H}, we wish to combine these values into a single parameter, $\hat{\delta}$ (not to be confused with the Kronecker delta, $\delta_{i,j}$).

\begin{figure*}
\includegraphics[width=1\textwidth]{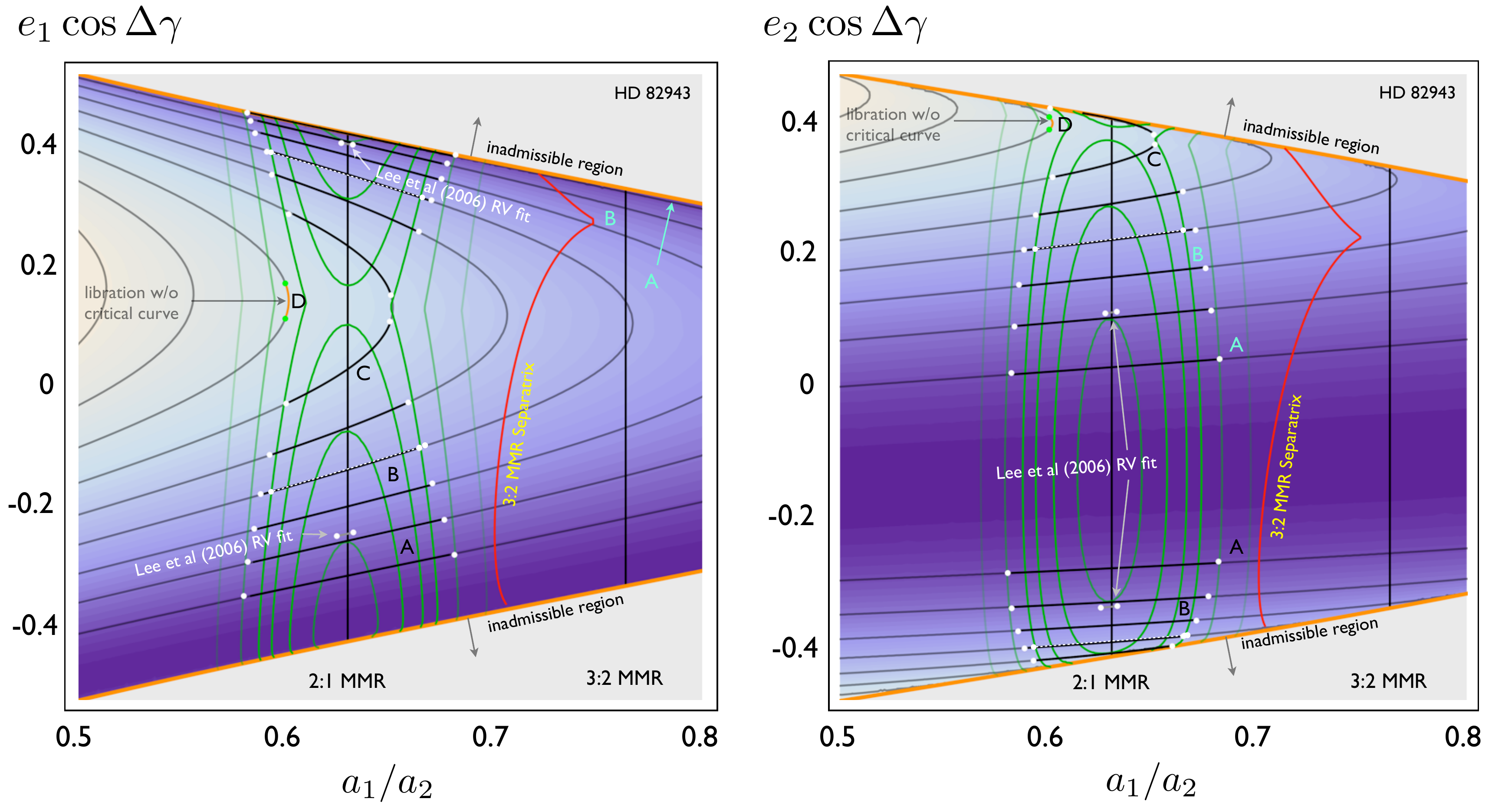}
\caption{A geometrical representation of the resonant dynamics. As in Figure (\ref{levelcurves}), the mass-ratio and angular momentum are chosen to be equivalent to that of the 2:1 resonant HD 82943 system (see Table 1). The two panels depict the surfaces of section of the dynamical evolution with respect to $\Delta \gamma$, taken at $\Delta \gamma = 0, \pm \pi$. As described in the text, showing both $e_1$ and $e_2$ is redundant and from the figures it can be understood that the two surfaces of section are a simple rotation away from each-other. In both panels, the permissible region is defined by a combination of the constants of motion $\mathcal{K}$ \& $\Omega$ and is in essence dictated by the angular momentum of a given system. Contours of $\Psi_2$ are shown as black lines and the background color is indicative of the value of $\Psi_2$. Namely, the dark blue region corresponds to a $\Psi_2$ minimum. For each plotted value of $\Psi_2$, the separatrix is mapped onto the figure using white dots. Maximal libration amplitudes for $\Psi_2$ levels that allow for the existence of a critical curve are traced along the corresponding contours with thick black lines. Maximal pseudo-resonant libration widths (i.e. those coresponding to $\Psi_2$ levels that do not allow for the existence of separatricies) are shown as thick orange lines and are bounded by yellow (caution: these points may appear green on some monitors and/or printers), rather than white dots. The $\Psi_2$ contours labeled A B C D parallel the dynamical portraits depicted in Figure (\ref{levelcurves}). The green vertical lines show constant energy levels. The approximate location of the separatrix of the neighboring 3:2 mean motion resonance is shown as a red curve. The \citep{2006ApJ...641.1178L} fit to the radial velocity data is shown with a gray line. The dashed white line depicts the initial condition of the system considered in section 3.2.}
\label{Poincare21}
\end{figure*}

Rescaling the actions and the Hamiltonian (while retaining the angles as before) by a constant factor $\eta$, such that 
\begin{eqnarray}
\label{scalethisbadboy}
\mathcal{K}' \rightarrow \mathcal{K}/\eta,  \  \nonumber \\
\Theta' \rightarrow \Theta/\eta, \ \   \nonumber \\
\Phi_1' \rightarrow \Phi_1/\eta, \nonumber \\
\Phi_2' \rightarrow \Phi_2/\eta,
\end{eqnarray}
we shall require that the constant in front of the $\Theta'^2$ term in (\ref{Hkeptheta}) be the same as the constant in (\ref{Hressuicidepolar}). In order for this transformation to remain canonical, we must also divide $\mathcal{H}$ by $\eta$. The expression for the scaling factor is:
\begin{equation}
\eta = \left( \frac{4(\alpha^2 + \beta^2)}{9([h]_1 (k-1)^2 + [h]_2 k^2)^2} \right)^{1/3}.
\end{equation}
Choosing to measure time in units of $3([h]_1 (k-1)^2 + [h]_2 k^2) /2$, which in turn allows us to again divide $\mathcal{H}$ by the same factor, we obtain an elegant expression for the scaled total Hamiltonian:
\begin{equation}
\mathcal{H} = \hat{\delta} \Theta' - \Theta'^2 - \sqrt{2 \Phi'_1} \cos(\phi_1 + \theta).
\end{equation}
Accordingly, the parameter $\hat{\delta}$ reads:
\begin{equation}
\hat{\delta} = \frac{2 [h]_1 (k-1) \mathcal{K}'}{([h]_1 (k-1)^2 + [h]_2 k^2)}.
\end{equation}

We now return to the issue of the final canonical transformation. The change of variables we are after is given by the following generating function of the second kind:
\begin{equation}
F_2 = \theta \Omega +  (\phi_1+\theta) \Psi_1 +  (\phi_2+\theta) \Psi_2.
\end{equation}
Taking the appropriate derivatives as above, we obtain our final action-angle variables:
\begin{eqnarray}
\label{finalactionangle}
\Psi_1 = \Phi'_1, \ \ \ \ \ \ \ \ \ \ \ \ \  \ \ \psi_1 = \phi_1 + \theta, \nonumber \\
\Psi_2 = \Phi'_2, \ \ \ \ \ \ \ \ \ \ \ \ \  \ \ \psi_2 = \phi_2 + \theta, \nonumber \\
\Omega = \Theta' - \Psi_1 -  \Psi_2, \ \ \ \ \ \ \ \ \ \omega = \theta.
\end{eqnarray}

Upon conversion to these variables, we arrive at an integrable one degree of freedom Hamiltonian:
\begin{eqnarray}
\label{Honedof}
\mathcal{H} = \hat{\delta} (\Omega + \Psi_1 + \Psi_2) - (\Omega + \Psi_1 + \Psi_2) ^2- \sqrt{2 \Psi_1} \cos(\psi_1).
\end{eqnarray}
Indeed, the newly defined angle $\omega$ is no longer present in the Hamiltonian (\ref{Honedof}), rendering $\Omega$ our last constant of motion. An examination of the expressions for $\Omega$ and $\mathcal{K}$ reveals that the total angular momentum of the (planar) system, $\mathcal{L}$, is given by 
\begin{equation}
\label{AMdef}
\mathcal{L} \simeq m_1 \sqrt{\mathcal{G} M a_1} (1 - \frac{e_1^2}{2}) + m_2 \sqrt{\mathcal{G} M a_2} (1 - \frac{e_2^2}{2}) = \eta(\Omega + \mathcal{K}')
\end{equation}
Furthermore, it is relevant to note that the planar angular momentum deficit, $\mathcal{A}$, which is conserved far away from mean motion resonances (i.e. in the secular domain - see \citet{1997A&A...317L..75L})
is conveniently given by 
\begin{equation}
\label{AMDdef}
\mathcal{A} = \Gamma_1 + \Gamma_2 = \eta(\Psi_1 + \Psi_2).
\end{equation}

At first glance, the transformation (\ref{finalactionangle}) may appear odd, because of the implicit choice to introduce the constant of motion $\Psi_2$ explicitly into the Hamiltonian. However, this selection is deliberate and will turn out to be useful in the next section, where it will be shown that the conservation of $\Psi_2$ is destroyed when higher-order interactions are taken into account (accordingly, the integrability of the Hamiltonian (\ref{Honedof}) is also compromised at higher order).

As already mentioned above, the Hamiltonian (\ref{Honedof}) is equivalent to the widely-discussed second fundamental model of resonance (\citet{1983CeMec..30..197H}, \citet{1986sate.conf..159P}, \citet{1999ssd..book.....M} and the references therein) and is therefore closely related to the pendulum model for resonance \citep{1976ARA&A..14..215P, 1973PhDT.........1Y}. In other words,  the dynamics of the unrestricted resonant problem exhibits qualitatively similar behavior to the broadly studied restricted problem, although the physical meanings of the involved variables are different.

An important distinction is that in the context of the Hamiltonian (\ref{Honedof}), the proximity to exact resonance is not given as a single parameter (as it is in the context of the restricted problem), but rather a combination of $\mathcal{K}, \Omega$ and $\Psi_2$. That said, it should be noted that $\mathcal{K}$ and $\Omega$ dictate the angular momentum surface on which the dynamics resides, while $\Psi_2$ is related to the initial conditions of the system confined on such a surface. Thus, it makes sense to treat $\Psi_2$ as an effective measure of proximity to exact resonance, given fundamental parameters of the system.

Level curves of the Hamiltonian for values of $\mathcal{K}$ and $\Omega$ that correspond to the HD 82943 entry in Table (1) \citep{2004A&A...415..391M} and various values of $\Psi_2$ are presented in Figure (\ref{levelcurves}). The curves on the panels correspond to different energy levels and are color-coded in order to highlight the distinct nature of the dynamics they entail. Specifically, the black and blue curves denote resonant and nonresonant trajectories respectively, while the gray curves on panels A, B and C denote critical curves that separate the resonant and nonresonant regions of the phase space. Note that the presence of the gray curves depends on the value of $\Psi_2$ that characterizes the plot. For instance, the orange curve on panel D describes libration in absence of a critical curve. Fixed points of the Hamiltonian are marked with black dots, where filled circles correspond to stable equilibria and the converse is true for open circles. A more detailed discussion of the motion described by these dynamical portraits and how they relate to the envisioned behavior of the orbits is presented below. 

\subsection{A Geometrical Representation of Resonant Dynamics}

\begin{figure*}
\includegraphics[width=1\textwidth]{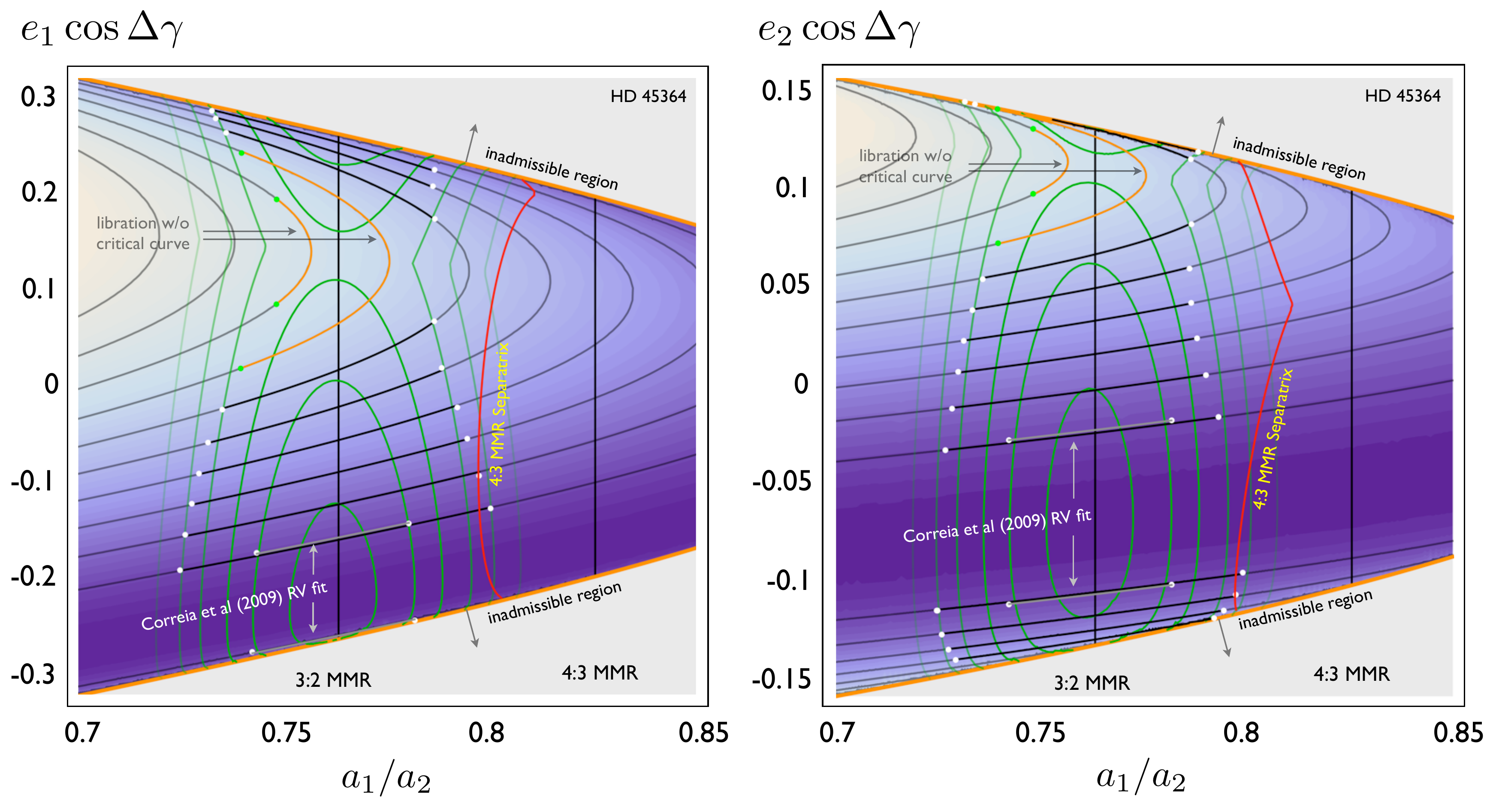}
\caption{Surfaces of section of the 3:2 resonant dynamics of the HD 45364 system.The meanings of the plotted curves are identical to those described in Figure (\ref{Poincare21}). However, the gray curve corresponds to the radial velocity solution of \citet{2009A&A...496..521C}. The same solution, at various levels of approximation is depicted in Figure (\ref{Correiacomp}).}
\label{Poincare32}
\end{figure*}

Armed with an integrable approximation to resonant motion, we may now take advantage of the various integrals identified above in order to formulate a geometrical representation of the orbital evolution. As a result of the numerous canonical transformations employed in the derivation of the Hamiltonian (\ref{Honedof}), the final variables (\ref{finalactionangle}) are rather serpentine. Consequently, here we shall opt to obtain the solutions as shown above, but subsequently work backwards through the transformations in order to represent the resonant behavior in terms of the Keplerian elements. 

We begin by defining the representative plane. As already mentioned in the discussion of transformation (\ref{transformone}), the actual values of the semi-major axes determine the timescale on which resonant perturbations occur, rather than the form of the interactions themselves \citep{1999ssd..book.....M}. We can therefore use the conservation of $\mathcal{K}$ to introduce the semi-major axis ratio, $a_1/a_2$ as our first independent variable. 

Keeping in mind that the problem we consider is effectively planar, it is natural to turn to the definition of the angular momentum, $\mathcal{L}$, for further development. Upon substitution of the definition of $\mathcal{K}$ and $a_1/a_2$ into equation (\ref{AMdef}), the conservation of $\mathcal{L}$ yields one of the eccentricities as a second independent variable. 

Because the variables ($\Psi,\psi$) implicitly originate from the definition of the vectors ($x,y$) given by equations (\ref{secularcart}), they encompass information about the planetary eccentricities as well as the difference in the longitudes of perihelia. Accordingly, a final requirement for the delineation of parameter space in question is a condition on the apsidal angles of the orbits. Clearly, the full time-evolution of the resonant dynamics requires a three-dimensional manifold, defined by ($a_1/a_2, e_1 \ \rm{or} \ e_2, \Delta \varpi$). However, a suitable representation of the dynamics can still be obtained by constructing a surface of section, choosing $\Delta \gamma = (0, \pm \pi)$ as an intersection plane\footnote{Note that the original perturbation Hamiltonians (\ref{Hkeporbel}) and (\ref{Hkepres}) are subject to D'Almbert rules, which physically imply invariance under rotation of the reference frame \citep{2002mcma.book.....M}. Consequently, only the difference of the perihelia, $\Delta \gamma$ is physically sensible, not the values of the angles themselves.}. Importantly, in our section, we shall not discriminate based on the direction of the trajectory's encounter with the $\Delta \gamma = (0, \pm \pi)$ plane. As will become clear below, this choice allows us to readily distinguish between apsidally librating and circulating orbits.

To summarize the above discussion, we choose to represent the resonant dynamics on a ($a_1/a_2, e \cos (\Delta \gamma) = \pm e$) surface of section. As expository examples, figures (\ref{Poincare21}) and (\ref{Poincare32}) depict such sections, roughly corresponding to the 2:1 resonant dynamics of the HD 82943 system \citep{2004A&A...415..391M,2006ApJ...641.1178L} and the 3:2 resonant dynamics of the HD 45364 system \citep{2009A&A...496..521C,2010A&A...510A...4R} respectively. The two panels in each figure show both, $e_1$ and $e_2$ for completeness, although as mentioned already, this is a redundancy, and upon examination it is clear that the two figures are a simple vertical rotation away from each other. 

Both of the planetary systems we use as examples here were detected by the radial velocity technique, and are comprised of giant planets around Sun-like stars. Notably, the masses of the planets are comparable ($m_2/m_1 \simeq 0.9$ for HD 82943 and $m_2/m_1 \simeq 3.5$ for HD 45364, see also Figure \ref{mratio}) preventing a description of the dynamics within the context of the restricted three-body problem. While it is firmly established that both of these systems are indeed resonant, some uncertainties exist in the orbital fits to the radial velocity data (see \citet{2006ApJ...641.1178L}, \citet{2010A&A...510A...4R}). We reiterate that for the purposes of this work, we shall only use these planetary pairs as illustrative examples, with little desire to quantify the exact nature of their dynamics. Furthermore, having picked a mass ratio and an angular momentum surface, we shall survey other parameters (e.g. orbital energy, values of $\Psi_2$) freely in order to epitomize an approximate yet global, rather than a precise but delimited picture of the dynamics. As a baseline, we shall adopt the (long-term stable) orbital solutions of  \citet{2006ApJ...641.1178L} for HD 82943 (specifically, Fit 2) and \citet{2009A&A...496..521C} for HD 45364 (listed in Table 1 for convenience), ignoring the possibility that improved fits to the data may yield somewhat different orbits.

Within the context of each section presented in Figures (\ref{Poincare21}) and (\ref{Poincare32}), an admissible region can be defined by the conservation of $\mathcal{L}$, along with the requirement that the eccentricities remain real. The admissible regions are delineated by bounding orange lines. Meanwhile, straight vertical lines depict the nominal semi-major axes of the shown resonances.

In addition to $\mathcal{K}$ and $\mathcal{L}$, the Hamiltonian (\ref{Honedof}) is characterized by conservation of $\Psi_2$. Consequently, within the admissible region, the dynamics must reside on contours of $\Psi_2$. These contours are shown as black lines with the background color indicating the values (dark blue stands for $\Psi_2 = 0$). Note that for the nominal value of $a_1/a_2$, each contour of $\Psi_2$ is intersected twice. For some contours (i.e. those below level A on Figure \ref{Poincare21}) both intersections take place at $\Delta \gamma = \pm \pi$. This means that the dynamics that reside on this $\Psi_2$ level are characterized by anti-aligned libration of the periapsis. For other levels of $\Psi_2$, (e.g level B of Figure \ref{Poincare21}), one intersection occurs for $\Delta \gamma = 0$ and one for $\Delta \gamma = \pm \pi$. This implies that the dynamics is characterized by circulation of the difference of the periapses. Finally, for higher levels of $\Psi_2$, (e.g. level C on the same figure), both intersections occur at $\Delta \gamma = 0$, implying that the dynamics is characterized by aligned apsidal libration. 

For each combination of $\mathcal{K}$, $\mathcal{L}$ and $\Psi_2$, there exists an energy level that separates librating and circulating orbits. Accordingly, such energy levels correspond to the maximal attainable libration widths and thereby define the resonant domain in parameter space \citep{2002mcma.book.....M}. Examples of such energy levels are shown as gray curves in panels A, B, C and as an orange curve in panel D of Figure (\ref{levelcurves}). If an energy level of this sort describes an orbit that passes through a hyperbolic fixed point, (shown as open circles on panels A, B and C of Figure \ref{levelcurves}), such an energy level is referred to as a separatrix or a critical curve. 

Strictly speaking, resonant orbits are exclusively those that reside within the croissant-shaped domain encompassed by the separatrix (see for example \citet{2012A&A...546A..71D}) e.g. the black curves in panels A, B, C in Figure (\ref{levelcurves}). Conversely, if regions of phase-space occupied by librating and circulating orbits are separated by a regular curve (such as the orange curve in panel D of Figure \ref{levelcurves}), none of the trajectories (even librating ones) are technically resonant, although numerous authors including ourselves (e.g. \citet{1902BuAsI..19..289P,1983CeMec..30..197H,2012arXiv1204.2791B}) have loosely used resonance and libration as synonyms. While in certain applications, this mixing of definitions does not pose significant problems, here, in interest of avoiding confusion, we shall refer to libration in absence of separatrix as a pseudo-resonance and retain the strict definition for true resonant motion. 

The boundary of the resonant domain is represented in Figures (\ref{Poincare21}) and (\ref{Poincare32}) by white points along contours of $\Psi_2$, while the pseudo-resonant domain is bounded by yellow points. In other words, regions confined by white points and shown as thick black curves, correspond to true separatricies such as those shown in panels A, B and C of Figure (\ref{levelcurves}). The regions confined by yellow points and shown as thick orange curves correspond to regular trajectories that separate libration and circulation. These points depict sections of the dynamics at $\psi_1 = \pi$. Accordingly, from Figure (\ref{levelcurves}), it is clear that resonant and pseudo-resonant orbits intersect the $\psi_1 = \pi$ line twice, as opposed to circulating orbits that intersect the $\psi_1 = \pi$ line only once\footnote{Inner circulators can exhibit an apparent libration of $\psi_1$ around 0 (see Figure \ref{levelcurves}). In this case, section is made with the $\psi_1 = 0$ axis on the left side of the stable equilibrium point.}.  

Note the deliberate parallel between curves labeled A, B, C, D in Figure \ref{Poincare21} and the panels in Figure \ref{levelcurves}). Indeed, the two figures represent the same dynamics, depicted in different spaces. The concave part of $\Psi_2$ contours, residing to the right of the resonant domain in Figures (\ref{Poincare21}) and (\ref{Poincare32}) (shown as thin black lines) can be identified as the inner circulation region shown in panels A, B, and C of Figure (\ref{levelcurves}), while the parameter space to the left of the resonant domain respectively corresponds to the outer circulation region. Of course, the inner circulation region disappears along with the separatrix. Consequently, the separatrix disappears on a level of $\Psi_2$ where the right hand side of the thick black curve reaches the rightmost extreme of the $\Psi_2$ contour. Because the two thick black curves on the upper and lower sides of the $\Psi_2$ contour are symmetric, the disappearance of the separatrix can be understood as taking place when the right-hand sides of two such curves join. Specifically, the curve labeled C in Figure \ref{Poincare21} is close to such a transition. 

\begin{figure}
\includegraphics[width=1\columnwidth]{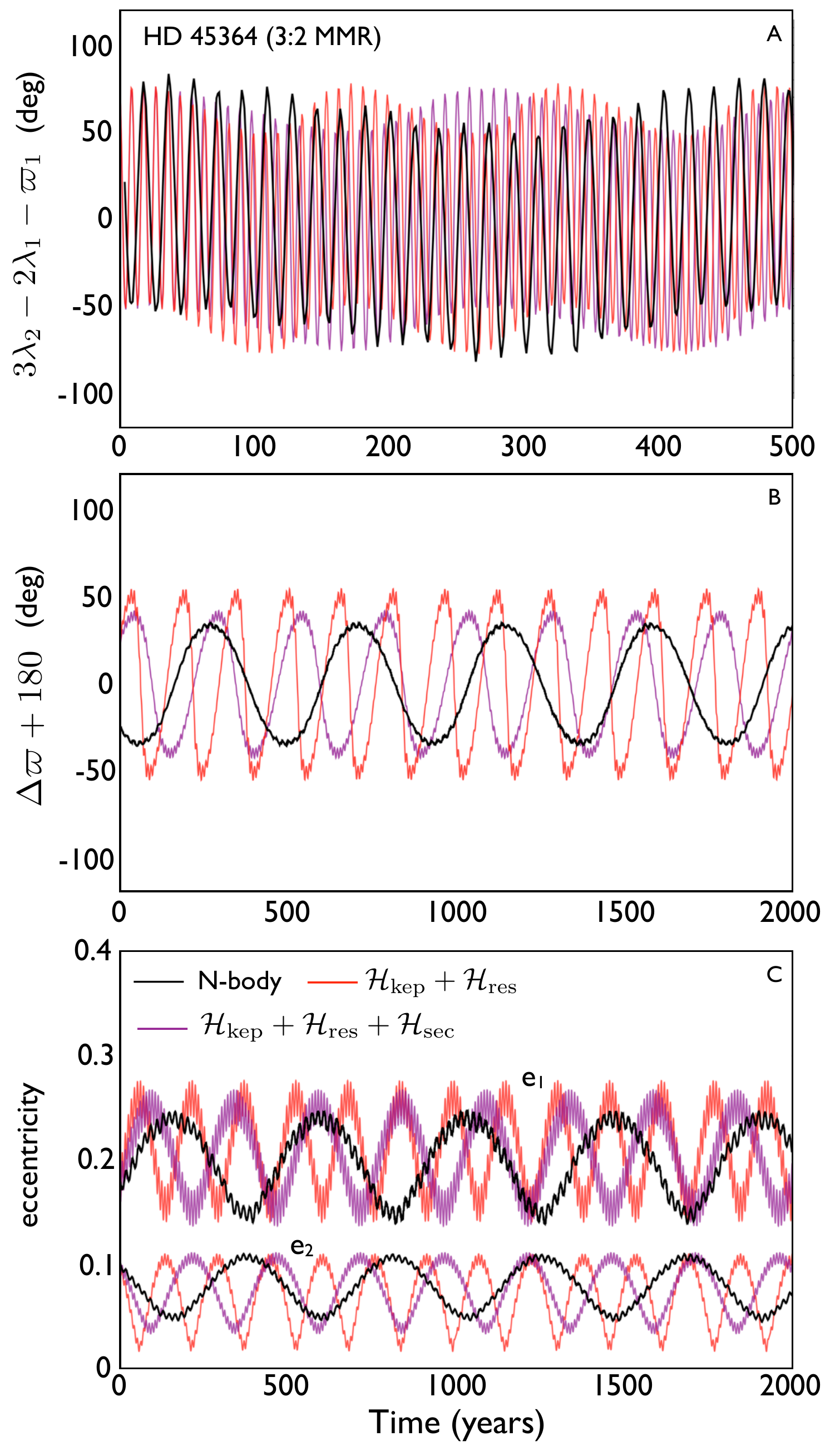}
\caption{The dynamical evolution of the \citet{2009A&A...496..521C} resonant fit at various levels of approximation. The black curve corresponds to the solution obtained with numerical N-body integration software (that is by direct integration of the Hamiltonian (\ref{Hbasic})). The red curves are given by the integrable Hamiltonian (\ref{Honedof}). The purple curves depict the dynamical evolution derived from the nonintegrable Hamiltonian (\ref{Htwodof}). Following \citet{2009A&A...496..521C}, panel A shows the evolution of resonant argument $3 \lambda_2 - 2 \lambda_1 - \varpi_1$. Panel B shows the evolution of the apsidal angle $\Delta \gamma$. Panel C shows the eccentricities as functions of time. The evolution of the second resonant argument, $3 \lambda_2 - 2 \lambda_1 - \varpi_2$, can be re-constructed by combining the evolution of $3 \lambda_2 - 2 \lambda_1 - \varpi_1$ with that of $\Delta \gamma$.}
\label{Correiacomp}
\end{figure}

Recall that together, $\mathcal{K}$, $\mathcal{L}$ and $\Psi_2$ constitute a measure of proximity to exact resonance. Indeed, once defined, we have all the ingredients to construct phase-space diagrams such as those shown in Figure (\ref{levelcurves}). Bearing in mind that each trajectory that resides on such a diagram is characterized by its energy level, we identify the value of $\mathcal{H}$ itself as the final geometrical constraint on the dynamics. Contours of $\mathcal{H}$, evaluated at $\Delta \gamma = (0,\pm \pi)$ and sectioned at $\psi_1 = \pi$, are shown as green lines on the surfaces of section. Importantly, the intersections of $\mathcal{H}$ contours with $\Psi_2$ contours depict the resonant libration width of a given configuration. Notice the inherent shape of $\mathcal{H}$: some contours are concave, while others are convex.

The best-fit orbital solutions of \citet{2006ApJ...641.1178L} and \citet{2009A&A...496..521C} are shown on the surfaces of section using white points connected by gray lines. Both solutions reside deep within the resonance signaling qualitative agreement of the integrable model with the true rendition of the dynamics. However, quantitatively, one should not expect the agreement between real resonant dynamics and the integrable approximations to be particularly good, because the convergence of the perturbation series employed here is questionable at best (especially at moderate eccentricities, provided that the orbits may intersect). Additionally, one should keep in mind the fact that thus far, we have neglected any terms in the disturbing function of order $e^2$ or greater, further spoiling the approximation \footnote{Note that Figure (\ref{Poincare21}) shows the HD82943 system in a state apsidal circulation whereas \citet{2006ApJ...641.1178L} observe anti-aligned libration for the same system. In addition to the errors inherent to the approximations involved, this difference may also stem from our usage of slightly different planetary masses.}.

Figure (\ref{Correiacomp}) illustrates a comparison between the dynamical evolution of the \citep{2009A&A...496..521C} fit obtained by numerical integration of Hamiltonian (\ref{Hbasic}) (shown as black lines) using the \textit{mercury6} integration package \citep{1999MNRAS.304..793C} and the analytical solution that arises from the approximate Hamiltonian (\ref{Honedof}) (shown as red lines). The panels A, B and C depict the time evolution of the resonant angle $\theta + \gamma_1 = 3 \lambda_2 - 2 \lambda_1 - \varpi_1$, the difference of the apsidal angles, $\Delta \varpi$ and the eccentricities respectively. Clearly, the characteristic frequencies of the oscillations differ by a factor of a few, however the amplitudes are well captured within the context of the approximate model.

\section{The Onset of Chaos}

\subsection{Overlap of Mean Motion Resonances}

The orbital architecture of small bodies in our solar system highlights the fact that resonances may exhibit both, regular and highly chaotic motion. In particular, while Neptune's external 3:2 and 2:1 mean motion resonances are densely populated with Kuiper belt objects \citep{2008ssbn.book..275M}, Jupiter's interior 2:1 and 3:1 resonances, that coincide with Kirkwood gaps of the asteroid belt are cleared out. The removal of resonant asteroids is now understood to be a result of chaotic diffusion that drives asteroids onto Mars-crossing orbits \citep{1985Icar...63..272W, 1987Icar...69..266H, 1990CeMDA..47...99H}. The same rationale is applicable to the unrestricted problem we address here. 

It is  well known that overlap among neighboring resonant domains gives rise to chaotic diffusion \citep{1969PhRv..188..416W,1979PhR....52..263C, 1980AJ.....85.1122W}. Consequently, the approximate (strictly periodic) model derived above is of virtually no use to the description of  energy levels that allow the corresponding orbits to penetrate neighboring resonances. In other words, the domain of applicability of the integrable model is in part determined by a given trajectory's proximity to a neighboring separatrix. 

Although there is no reason why all separatricies should lie on the same energy level (in fact they don't), it can be seen in Figures (\ref{Poincare21}) and (\ref{Poincare32}) that the resonant and pseudo-resonant domains are approximately bounded by the transition between concave and convex energy levels. This similarity can be taken advantage of, to map the approximate locations of the neighboring resonances. This portrayal of the onset of chaos is by no means intended to be precise and is strictly speaking heuristic since the separatricies are obtained by sectioning each resonant dynamics relative to different critical angles, even if they lie on the same angular momentum surface. However, we do not view this as a significant drawback, since the Chirikov resonance overlap criterion is in itself an approximation that neglects the coupling of the resonances and the resulting deformation of their shape, as well as the generation of higher order secondary resonances (that act to expand the size of the chaotic zone). 

As an example, consider the level of $\Psi_2$ adjacent to the \citet{2009A&A...496..521C} orbital solution shown in Figure (\ref{Poincare32}). While it was shown in the previous section that the orbital fit itself (roughly corresponding by the second inner most energy level) is moderately well represented by the analytical Hamiltonian (\ref{Honedof}) (see Figure \ref{Correiacomp}), we can anticipate that the same will not be true of the solution characterized by the second outermost energy level, since the orbit resides in close proximity to the separatricies of the 3:2 and the 4:3 resonances. 

A straight forward way to account for the effects of both, the 3:2 and the 4:3 mean motion resonances is to construct a Hamiltonian of the form 
\begin{equation}
\label{MMRoverlap}
\mathcal{H} = \mathcal{H}_{\rm{kep}}^{(0)} + \mathcal{H}_{\rm{res}}^{(3:2)} + \mathcal{H}_{\rm{res}}^{(4:3)} + \mathcal{O} (e^2 , i^{2}),
\end{equation}
where the Keplerian term is given by equation (\ref{Hkeplambda}) and the two resonant contributions each take the form of equation (\ref{Hrespoincare}), with $k = 3,4$ and appropriately chosen coefficients, $f_{\rm{res}}$ (note that choosing to not expand the Keplerian Hamiltonian around any nominal resonant location further contributes to the nonlinearity of the system and acts to expand the chaotic zone). Such a Hamiltonian possesses four degrees of freedom and four harmonics hindering further simplification. As a result, we integrate the equations of motion that result from the Hamiltonian (\ref{MMRoverlap}) using conventional numerical methods. 

\begin{figure}
\includegraphics[width=1\columnwidth]{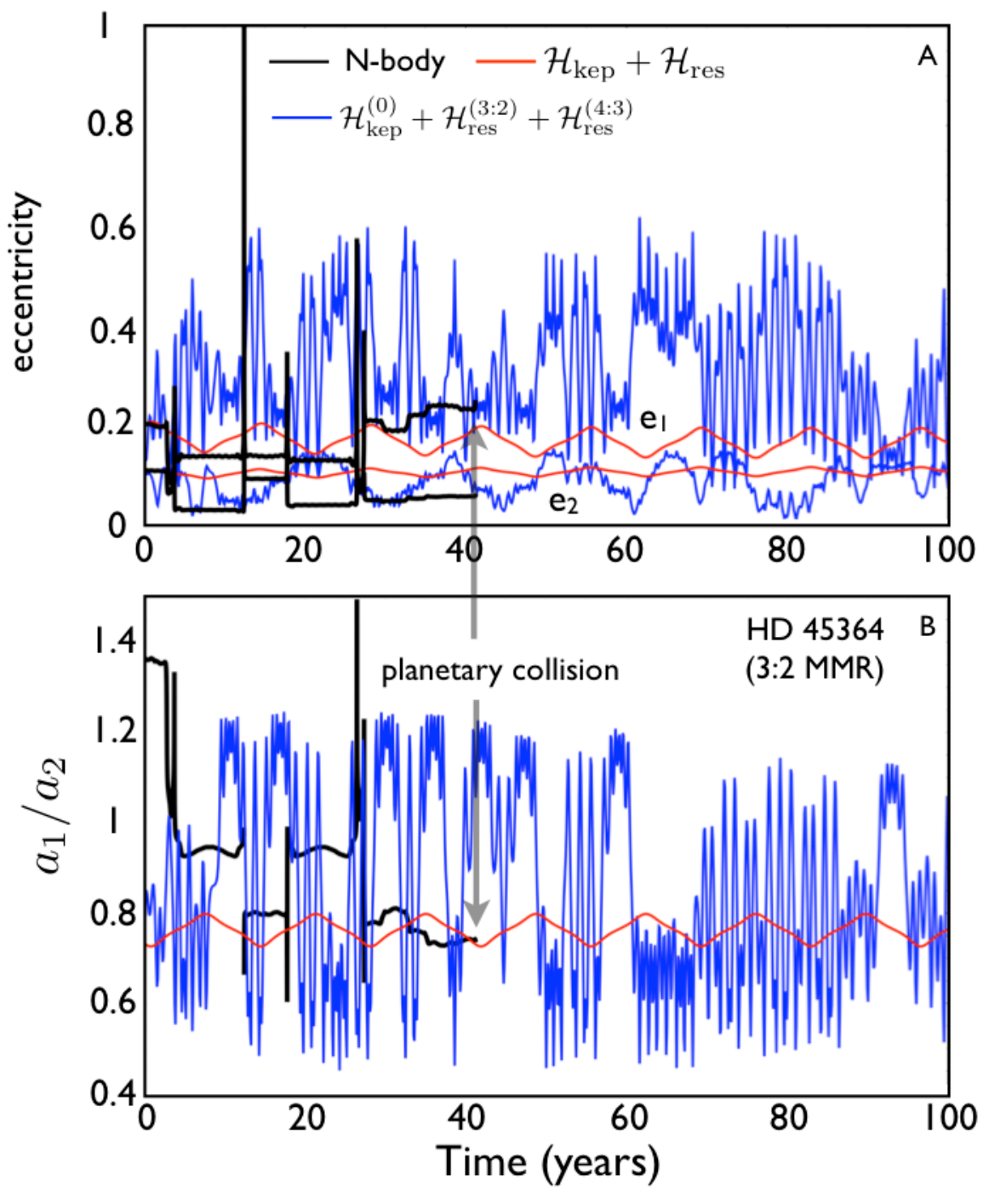}
\caption{Chaotic evolution of a 3:2 resonant system at differing levels of approximation. Panels A and B show the orbital eccentricities and the semi-major axis ratio respectively. The initial conditions of the described system correspond to the same level of $\Psi_2$, as that of the \citet{2009A&A...496..521C} fit. However, the energy level is chosen such that the motion overlaps the 4:3 resonance. Specifically, the second outermost energy level, plotted in Figure (\ref{Poincare32}) is considered. As in Figure (\ref{Correiacomp}), the black curve corresponds to the result of a numerical N-body integration. Within the context of the N-body solution, the planets reverse order shortly after the beginning of the integration and subsequently suffer a mutual collision after $\sim$40 years of dynamical evolution. The blue curve represents the results obtained by numerical integration of the perturbative Hamiltonian (\ref{MMRoverlap}), that contains both resonant arguments. The inadequacy of the integrable approximation (\ref{Honedof}) (plotted in red) in a chaotic domain can easily be seen.} 
\label{Correiachaos}
\end{figure}

The resulting solution exhibits rapid dynamical chaos, as is made evident by the eccentricity and semi-major axis ratio evolution shown in figure (\ref{Correiachaos}) with blue lines. Indeed, the timescale for the onset of irregularity is comparable to the orbital timescale. We have repeated the numerical experiment with an N-body simulation as above and confirmed the fully chaotic nature of the configuration in question. Specifically, within the context of the N-body simulation, the stochastic evolution comes to a rapid end $\sim 40$ years into the integration, when the planets collide. The N-body results are shown in the figure with black lines. Meanwhile, the analytical model given by the Hamiltonian (\ref{Honedof}) predicts regular oscillations for the same configuration, as can be gauged from the red lines shown in the figure. 

Here we have chosen a somewhat extreme example to demonstrate planetary chaos. However, this exercise highlights the dangers and the associated care that must be taken during application of the simple model described in the previous section.

\subsection{Secular modulation of Resonant Motion}

Let us now turn our attention to a region of the resonant domain that is well-separated from the neighboring mean motion resonances. Over a sufficiently short period of time (which is related to the timescale on which resonant interactions exchange energy and angular momentum between the orbits), the Hamiltonian (\ref{Honedof}) provides a suitable approximation to the motion. However, if we wish to characterize the behavior of the system over a longer (secular) timescale, we are forced to retain additional terms in the disturbing function \citep{1996CeMDA..64..115L}. This is due to the fact that characteristic resonant frequencies are proportional to $\propto \sqrt{m/M}$ while secular frequencies are proportional to $\propto m/M$, giving rise to an inherent separation of timescales between the integrable Hamiltonian and the secular correction \citep{1990CeMDA..47...99H}.

As such, we extend our perturbation series to account for second-order secular coupling of the orbits\footnote{For $k:1$ type resonances, asymmetric resonant librations are possible \citep{1994CeMDA..60..225B,2012arXiv1211.3078K}. Consequently, for certain combinations of parameters, the phase-space portrait of the 2:1 mean motion resonance may be topologically different from that shown in Figure (\ref{levelcurves}).}:
\begin{eqnarray}
\label{Hseckep}
\mathcal{H} \simeq \mathcal{H}_{\rm{kep}} + \mathcal{H}_{\rm{res}} + \mathcal{H}_{\rm{sec}} + \mathcal{O} (e^2 , i^{2}),
\end{eqnarray}
where in terms of Keplerian orbital elements \citep{1999ssd..book.....M},
\begin{equation}
\mathcal{H}_{\rm{sec}} = -\frac{\mathcal{G} m_1 m_2}{a_2} ( f_{\rm{sec}}^{(1)} ( e_1^2 + e_2^2) + f_{\rm{sec}}^{(2)} e_1 e_2 \cos(\varpi_1 - \varpi_2)).
\end{equation}
As before, we shall evaluate $\mathcal{H}_{\rm{sec}}$ at nominal semi-major axes, rendering $f_{\rm{sec}}$ constant coefficients that depend on the semi-major axis ratios only. It should be noted that $\mathcal{H}_{\rm{sec}}$ does not provide the only secular contribution to the dynamics at second order in $e$. Resonant terms at second order in $e$, once averaged over a libration or a circulation cycle of $\psi_1$, also give rise to pure secular terms that can be as large as those given in (\ref{Hseckep}). Here, we opt to discard such terms for the sake of simplicity, especially given that our aim is merely to demonstrate the qualitative impact of secular terms (that is the generation of chaos) on the integrable approximation developed above. 

\begin{figure}
\includegraphics[width=1\columnwidth]{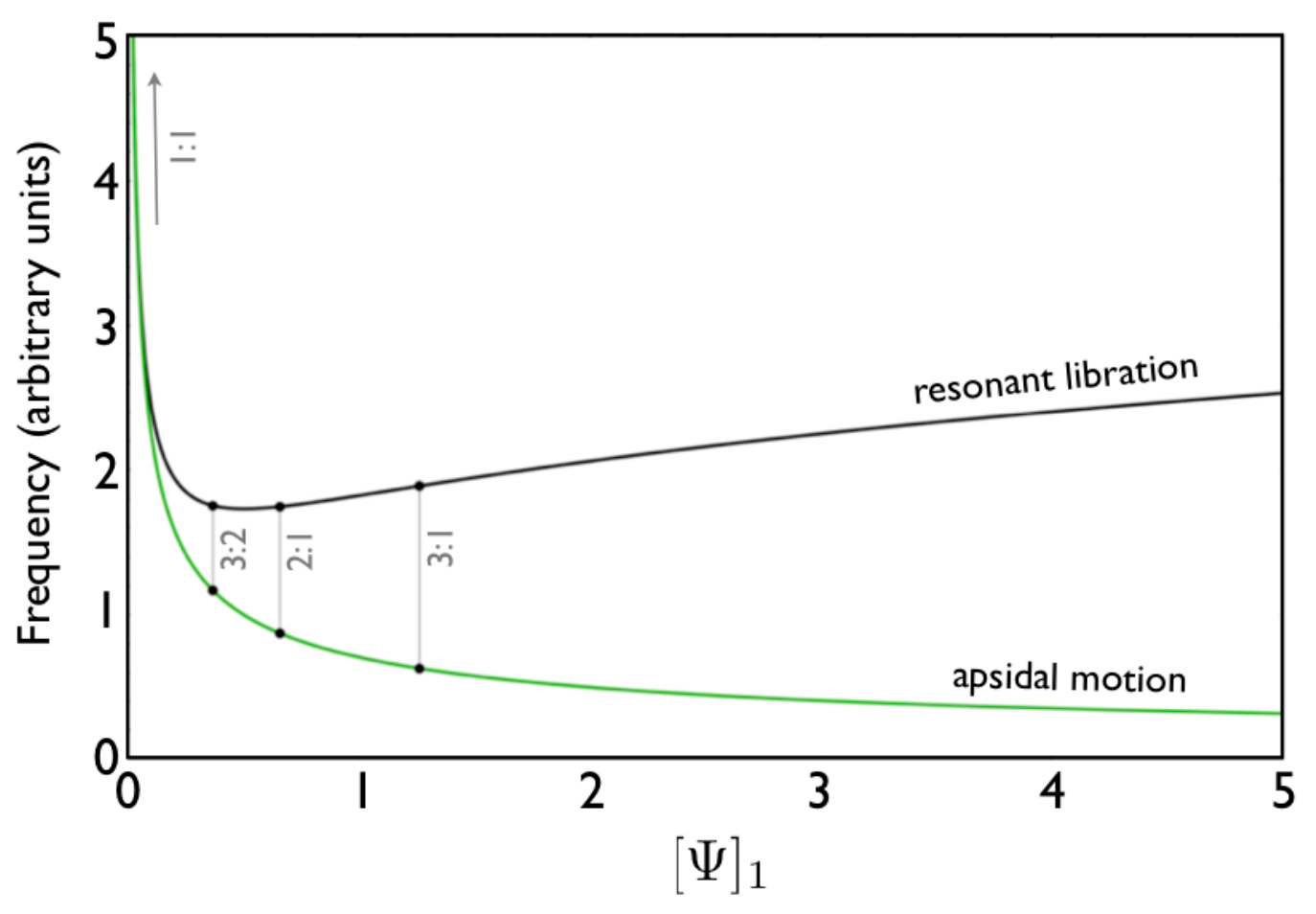}
\caption{The resonant libration frequency and the apsidal motion frequency as a function of $[\Psi]_1\propto e^2$ (see equations (\ref{suicide}) and (\ref{scalethisbadboy})), as calculated within the framework of Hamiltonian (\ref{Honedof}). For reference, the 3:2, 2:1 and the 3:1 secondary resonances are labeled. Note that the condition for appearance of secondary resonances implies low eccentricities.}
\label{secondaryres}
\end{figure}

Following the same procedure outlined in the derivation of the resonant Hamiltonian, we first revert to Poincare action-angle coordinates. The secular Hamiltonian now takes the form: 
\begin{equation}
\label{Hsecpoincare}
\mathcal{H}_{\rm{sec}} = -2 \mu \Gamma_1 - 2 \sigma \Gamma_2 - 2 \nu \sqrt{\Gamma_1 \Gamma_2} \cos(\gamma_2 - \gamma_2),
\end{equation}
where
\begin{eqnarray}
\label{seccoeff}
\mu =  \frac{\mathcal{G}^2 M m_1 m_2^3}{[\Lambda]_2^2} \frac{ f_{\rm{sec}}^{(1)}}{[\Lambda]_{1}}, \ \ \ \ \ \ \ \ \ \   \nonumber \\
\sigma =  \frac{\mathcal{G}^2 M m_1 m_2^3}{[\Lambda]_2^2} \frac{ f_{\rm{sec}}^{(1)}}{[\Lambda]_{2}},  \ \ \ \ \ \ \ \ \ \  \nonumber \\ 
\nu =  \frac{\mathcal{G}^2 M m_1 m_2^3}{[\Lambda]_2^2} \frac{ f_{\rm{sec}}^{(2)}}{\sqrt{[\Lambda]_{1} [\Lambda]_{2}}}.
\end{eqnarray}

It is noteworthy that Hamiltonian (\ref{Hsecpoincare}) depends only on a single harmonic and can thus be easily transformed into a one degree of freedom Hamiltonian, recognizing the angular momentum deficit, $\mathcal{A}$, as a secular constant of motion. Indeed, in isolation, $\mathcal{H}_{\rm{sec}}$ is integrable and the solution is referred to as the Laplace-Lagrange secular theory \citep{1999ssd..book.....M}. Upon employing the transformation to eccentricity vectors given by equations (\ref{secularcart}), the linear nature of the equations of motion that arise from $\mathcal{H}_{\rm{sec}}$ becomes apparent:
\begin{equation}
\label{Hsecpoincart}
\mathcal{H}_{\rm{sec}} = - \mu (x_1^2 + y_1^2) - \sigma (x_2^2 + y_2^2) - \nu (x_1 x_2 + y_1 y_2). 
\end{equation}

An application of the canonical rotation transformation (\ref{suicide}) converts $\mathcal{H}_{\rm{sec}}$ into a more cumbersome form:
\begin{eqnarray}
\mathcal{H}_{\rm{sec}} = ({\alpha ^2+\beta ^2})^{-1} (- (u_1^2 + v_1^2) (\alpha ^2 \mu +\alpha  \beta  \nu +\beta ^2 \sigma ) \nonumber \\ 
- (u_2^2+v_2^2)  (\alpha ^2 \sigma -\alpha  \beta  \nu +\beta ^2 \mu ) \nonumber \\ 
+ (u_1 u_2 + v_1 v_2) (\alpha ^2  \nu +2 \alpha  \beta  (\sigma -\mu )-\beta ^2 \nu ) ).
\end{eqnarray}
Finally, combining transformations (\ref{suicidepolar}) \& (\ref{finalactionangle}), and rescaling the time as above, we can express the full Hamiltonian as:
\begin{eqnarray}
\label{Htwodof1}
\mathcal{H} = \hat{\delta} (\Omega + \Psi_1 + \Psi_2) - (\Omega + \Psi_1 + \Psi_2) ^2 - \sqrt{2 \Psi_1} \cos(\psi_1) \nonumber \\
 - 4 (3 \eta (\alpha^2 + \beta^2)([h]_1 (k-1)^2 + [h]_2 k^2 ))^{-1} \nonumber \\
 \times  ( \alpha^2 (\mu \Psi_1 + \sigma \Psi_2)  + \beta^2 (\sigma \Psi_1 + \mu \Psi_2) +  \alpha \beta \nu (\Psi_1 - \Psi_2)  \nonumber \\
- (\alpha^2 \nu - \beta^2 \nu + 2 \alpha \beta (\sigma - \mu)) \sqrt{\Psi_1 \Psi_2} \cos (\psi_1 - \psi_2) ). 
\end{eqnarray}

The Hamiltonian (\ref{Htwodof1}) is characterized by two degrees of freedom, and as will become apparent shortly, exhibits chaotic motion. This implies that no canonical transformation can be found to identify additional constants of motion. However, prior to working with equation (\ref{Htwodof1}), it is worthwhile to examine the timescales on which the two degrees of freedom evolve, and identify the relevant regimes of motion, corresponding to commensurability and separation between the characteristic frequencies. Let us fist examine the conditions for commensurability and the generation of secondary resonances.

\subsubsection{Secondary Resonances}

In the framework of the unrestricted resonance problem, the numerical simulations of \citet{2008MNRAS.387..747M} showed that at very low eccentricities, secular and resonant angles can evolve on comparable timescales, giving rise to secondary resonances. With an integrable approximation to resonant motion in hand, we can examine the criteria for the appearance of secondary resonances analytically. More specifically, we shall aim to find conditions under which the period of resonant libration is close to a low-order integer ratio with the apsidal period.

To estimate the former, we expand Hamiltonian (\ref{Honedof}) as Taylor series in ($\Psi_1,\psi_1$) to second order, around the resonant equilibrium point, ($[\Psi]_1,\pi$). Dropping constant terms, and defining the variables
\begin{eqnarray}
\label{equilibvars}
\bar{\Psi}_1=\Psi_1 - [\Psi]_1 \ \ \ \ \ \ \ \ \ \ \ \ \bar{\psi}_1=\psi_1 - \pi,
\end{eqnarray}
we have:
\begin{eqnarray}
\label{Hsho1}
\mathcal{H} &=& -\frac{\sqrt{2 [\Psi]_1}}{2} \bar{\psi}_1^2 + \left(\hat{\delta} -2 \left([\Psi]_1 + \Psi_2 + \Omega \right) + \frac{1}{\sqrt{2 [\Psi]_1}} \right) \bar{\Psi}_1 \nonumber \\
&-& \left(1 + \frac{\sqrt{2}}{8\sqrt{([\Psi]_1)^3}} \right) \bar{\Psi}_1^2.
\end{eqnarray}
As long as the barred quantities remain small (that is, the system does not deviate away from equilibrium much), this simplification directly implies nearly-constant eccentricities and apsidal anti-alignment of the orbits i.e. $\Delta \gamma \simeq \pi$. Furthermore, because we are expanding the Hamiltonian around a fixed point, to linear order, $d\bar{\psi}_1/dt = 0$, meaning 
\begin{equation}
\label{nomres}
\hat{\delta} -2 \left([\Psi]_1 + \Psi_2 + \Omega \right) + \frac{1}{\sqrt{2 [\Psi]_1}} = 0.
\end{equation}
This expression automatically defines the nominal value of the action $[\Psi]_1$ for a given combination of $\hat{\delta}$, $\Omega$ and $\Psi_2$, while further simplifying the Hamiltonian (\ref{Hsho1}), as now only the quadratic terms remain.

Finally, after applying the transformation 
\begin{eqnarray}
\label{sqrttrans}
\tilde{\Psi}_1= \bar{\Psi}_1 \left ( \frac{1+4 \sqrt{2 \left([\Psi]_1\right)^3}}{4 \left([\Psi]_1\right)^2} \right)^{\frac{1}{4}}  \ \ \ \tilde{\psi}_1= \bar{\psi}_1 \left ( \frac{4 \left([\Psi]_1\right)^2}{1+4 \sqrt{2 \left([\Psi]_1\right)^3}} \right)^{\frac{1}{4}},
\end{eqnarray}

Hamiltonian (\ref{Hsho1}) reduces to that of a harmonic oscillator:
\begin{equation}
\label{secres1}
\mathcal{H} =  -  \frac{1}{2} \sqrt{\frac{2^{1/2} +8 \left([\Psi]_1 \right)^{3/2}}{2^{3/2} \left([\Psi]_1 \right)}} \left(\tilde{\psi}_1^2 + \tilde{\Psi}_1^2 \right) = \frac{\varphi_{\rm{res}}}{2} \left(\tilde{\psi}_1^2 + \tilde{\Psi}_1^2 \right).
\end{equation}
where $\varphi_{\rm{res}}$ is immediately identified as the resonant libration frequency.

By working back through the canonical transformations outlined in the previous section, it can be easily shown that because $d\bar{\psi}_1/dt = 0$ to leading order, the average apsidal frequency coincides with that of the cyclic angle $\psi_2$. An application of Hamilton's equations yields
\begin{equation}
\label{dpsi2dt}
\frac{d \psi_2}{d t} = \hat{\delta} -2 \left([\Psi]_1 + \Psi_2 + \Omega \right) = - \frac{1}{\sqrt{2 [\Psi]_1}},
\end{equation}
where the latter equality follows from equation (\ref{nomres}).

Equating the two frequencies, we find that the 1:1 secondary resonance only exists in the unphysical limit of $[\Psi]_1 \rightarrow 0$, which corresponds to null eccentricities. However, higher order secondary resonances are indeed permitted at small values of $[\Psi]_1$, in agreement with the work of \citet{2008MNRAS.387..747M} (see also \citet{1993Icar..103...99M}). Figure (\ref{secondaryres}) shows the two frequencies as a function of $[\Psi]_1$, and the 3:2, 2:1 and the 3:1 secondary resonances are labeled for reference. 

Having dropped the secular terms from the Hamiltonian (\ref{Hsho1}), we have implicitly limited the scope of the above calculations to systems where $\psi_2$ circulates. It is however important to note that upon inclusion of secular terms, libration of $\psi_2$ is possible within a limited range of parameter space, rendering the above calculation inapplicable. Such configurations will be discussed in section 5.

\subsubsection{Adiabatic Evolution}

\begin{figure*}
\includegraphics[width=1\textwidth]{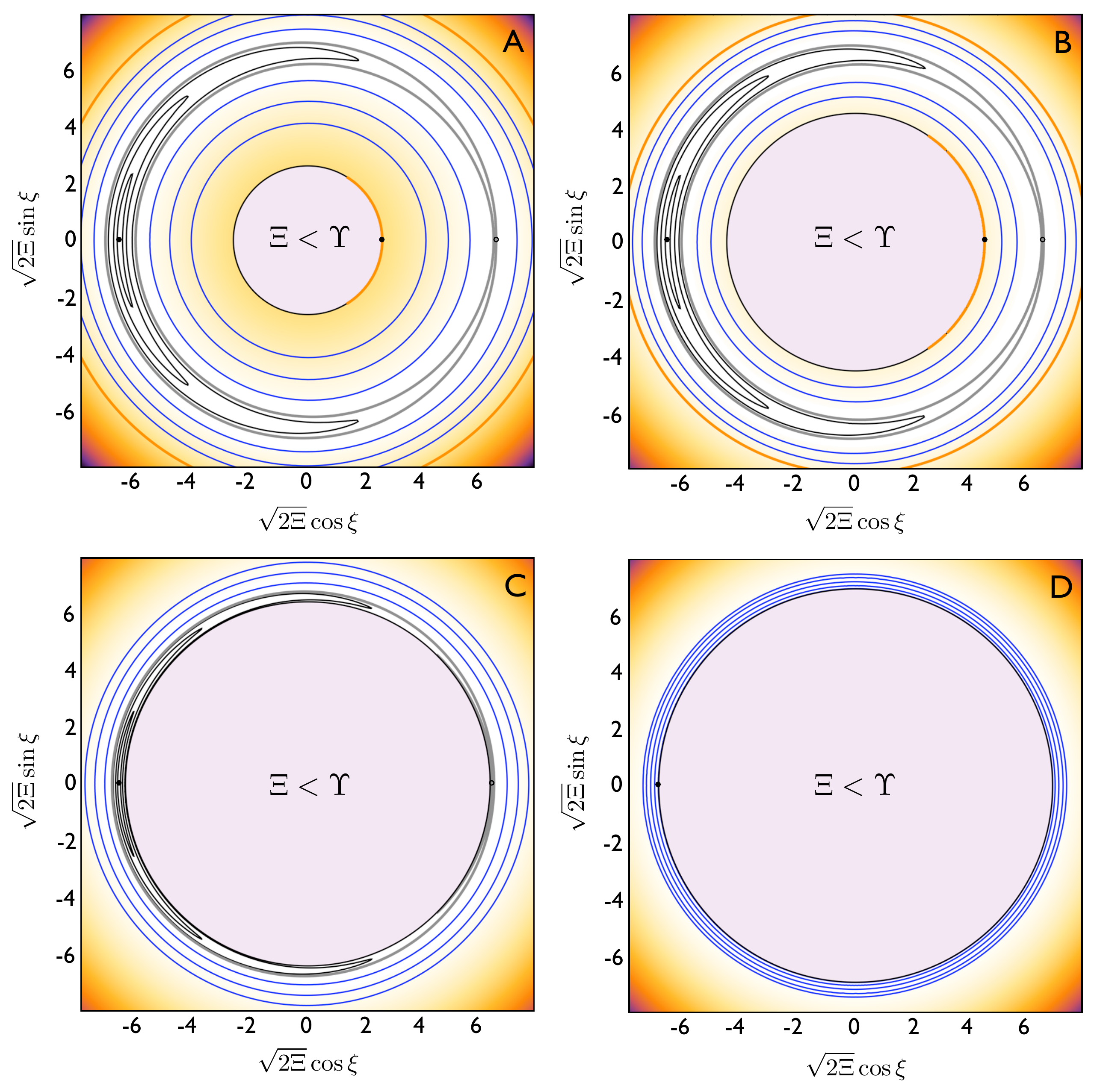}
\caption{Level curves of a frozen system given by the Hamiltonian (\ref{Htwodof}). The energy levels are plotted in the ($\Xi,\xi$) plane, freezing the second (slow) degree of freedom at $\Upsilon = \rm{const.}, \upsilon = 0$. Panels A B C D are characterized by the same values of $\Upsilon$ (that is, $\Psi_2$) as those shown in Figure (\ref{levelcurves}). The inadmissible region that arises once the Hamiltonian is formulated in terms of action-angle variables (\ref{amdactionangle}), is shown in light purple. Otherwise, the color scheme of the curves is the same as that shown in Figure (\ref{levelcurves}).}
\label{levelcurves2}
\end{figure*}

We now turn our attention away from the characteristic domain of secondary resonances, and towards the parameter regime that is more typical of the exoplanetary systems discussed in the previous section. As shown in Figure (\ref{secondaryres}), low-order secondary resonances are not possible if $[\Psi]_1$ is sufficiently large. In this case, the libration frequency of $\psi_1$ is much higher than the circulation frequency of the secular angle, $\psi_1 - \psi_2$. It is therefore sensible to transform the variables accordingly and employ the separation of timescales between the two degrees of freedom to our advantage \citep{1982amdc.proc..153H}.

The transformation we seek is given by the generating function 
\begin{equation}
F_2 = \psi_1 \Xi + (\psi_1 - \psi_2) \Upsilon,
\end{equation}
which yields the variables
\begin{eqnarray}
\label{amdactionangle}
\Xi = \Psi_1+ \Psi_2, \ \ \ \ \ \ \ \ \ \ \ \ \ \ \xi = \psi_1, \nonumber \\
\Upsilon = \Psi_2, \ \ \ \ \ \ \ \ \ \ \ \ \  \ \ \upsilon = \psi_1 - \psi_2.
\end{eqnarray}
The new action-angle variables are actually somewhat more intuitive than the previous. Specifically, as can be understood from equation (\ref{AMDdef}), $\Xi = \mathcal{A}/\eta$ is simply the re-scaled angular momentum deficit. Meanwhile, a modulation of the action, $\Upsilon$, changes the $\Psi_2$ contour on which the resonant dynamics resides in surfaces of section (\ref{Poincare21}) and (\ref{Poincare32}). In terms of the new variables, the Hamiltonian reads:
\begin{eqnarray}
\label{Htwodof}
\mathcal{H} = \hat{\delta} (\Omega + \Xi) - (\Omega + \Xi) ^2 - \sqrt{1- \Upsilon/\Xi} \sqrt{2 \Xi} \cos(\xi) \nonumber \\
- 4 (3 \eta (\alpha^2 + \beta^2)([h]_1 (k-1)^2 + [h]_2 k^2 ))^{-1} \nonumber \\
 ( \alpha^2 (\mu \Xi + (\sigma - \mu) \Upsilon) + \beta^2 (\sigma \Xi + (\mu - \sigma) \Upsilon) \nonumber \\
+  \alpha \beta \nu (\Xi - 2 \Upsilon) - (\alpha^2 \nu - \beta^2 \nu + 2 \alpha \beta (\sigma - \mu))  \nonumber \\
\times \sqrt{\Xi/\Upsilon -1 } \Upsilon \cos (\upsilon) ). 
\end{eqnarray}

Before considering an example that highlights the onset of chaos through secular modulation, let us reflect on the somewhat satisfactory agreement between the N-body simulation and the analytical treatment of the \citep{2009A&A...496..521C} orbital solution shown in Figure (\ref{Correiacomp}). A numerical solution of the equations of motion that arises from the Hamiltonian (\ref{Htwodof}) (using the initial conditions listed in Table 1) is shown with purple lines in Figure (\ref{Correiacomp}). This solution demonstrates that rather than introducing chaos, the addition of secular terms (unsurprisingly) improves the agreement between the perturbative treatment of the dynamics and the N-body simulation. Specifically, both the amplitude and frequency of oscillations in the apsidal angle $\Delta \gamma$ and the eccentricities are decreased compared to the analytical results stemming from the Hamiltonian (\ref{Honedof}), better matching the N-body calculations. Indeed, the introduction of higher-order terms does not render the entire phase-space chaotic.

The dynamical portrait of a two degrees of freedom system, cannot be represented visually in a simple fashion. However, it is still instructive to visualize the behavior of one of the degrees of freedom by freezing the evolution of the second degree of freedom. In particular, here we choose to set $\Upsilon = \rm{const.}, \upsilon = 0$. This is especially relevant to the dynamics described by the Hamiltonian (\ref{Htwodof}) because the evolution timescales of the two degrees of freedom are well-separated.

Maintaining a parallel with the discussion of the previous section, Figure (\ref{levelcurves2}) shows surfaces of section of the level curves of the Hamiltonian (\ref{Htwodof}), for the same values of $\Upsilon = \Psi_2$ as those shown in Figure (\ref{levelcurves}). As before, black, blue and gray curves denote resonant orbits, nonresonant orbits, and separatricies respectively. Pseudo-resonant orbits are marked as orange lines and are shown in panels A and B. As expected, these pseudo-resonant trajectories do not circle the center of the figure and therefore imply libration. Filled black dots denote stable equilibria while open circles mark unstable fixed points. Although Figures (\ref{levelcurves2}) and (\ref{levelcurves}) show essentially the same dynamical portraits, it can be argued that visualization in terms of the variables (\ref{amdactionangle}) is more instructive. 

Most importantly, the phase-space portrait retains the same location of the separatrix for all values of $\Upsilon$, at the expense of introducing an inadmissible region, marked by a light purple circle centered on the origin. The inadmissible region itself is defined by the condition $\Xi \leqslant \Upsilon$. Recalling that $\Xi$ is related to the angular momentum deficit, the physical interpretation of the inadmissible region is simply the requirement that the eccentricities never acquire a complex component: $\Im(e) = 0$. An equivalent interpretation of the boundary of the inadmissible region is that it represents a stretched out origin of the panels in Figure (\ref{levelcurves}) and thus corresponds to $e_1 = e_2 = 0$. 

An advantage of this representation is that the disappearance of the separatrix can be easily understood to be a result of the changes in $\Upsilon$. Indeed as the value of $\Upsilon$ grows from panel C to panel D, the separatrix is engulfed by the inadmissible region, leaving only nonresonant trajectories to fill the phase-space. 

With this interpretation in mind, it can be intuitively understood why the introduction of secular terms into our model can give rise to chaotic motion. Namely, while $\Upsilon$ is a constant of motion in the context of the Hamiltonain (\ref{Honedof}), it ceases to be constant with the introduction of secular terms. Referring back to Figures (\ref{Poincare21}) and (\ref{Poincare32}), the modulation in $\Upsilon$ can be visualized as a vertical translation across contours of $\Psi_2$, while confined to a particular energy level, denoted by green lines. 

\begin{figure}
\includegraphics[width=1\columnwidth]{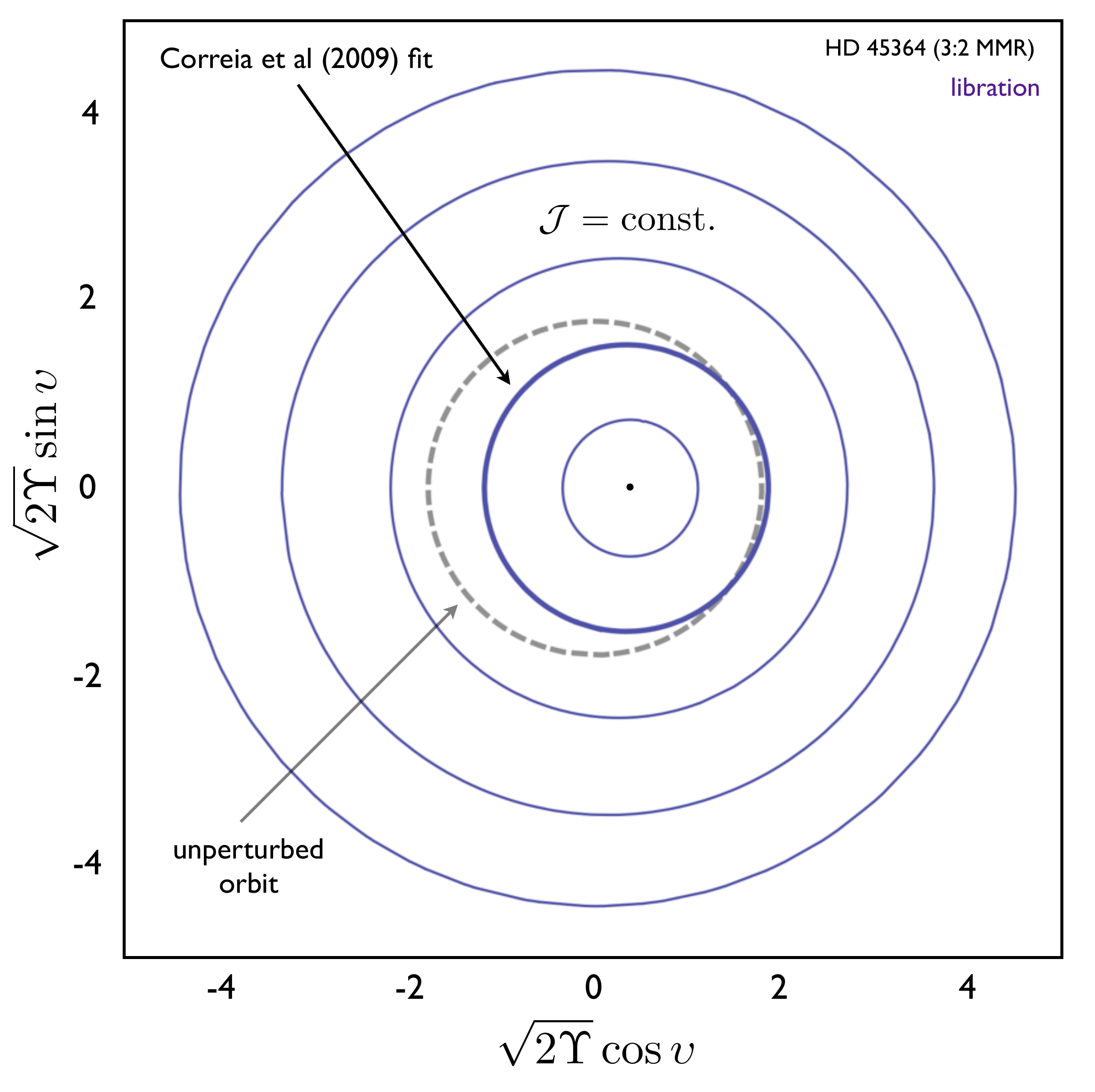}
\caption{A Poincare surface of section corresponding to the energy and angular momentum levels of \citet{2009A&A...496..521C}. Various starting values of $\Upsilon$ are considered and the solutions arising from the Hamiltonian (\ref{Htwodof}) are plotted as solid lines. The thick curve corresponds to the evolution originating from the actual \citep{2009A&A...496..521C} orbital fit. For reference, the trivial ($\Upsilon,\upsilon$) evolution of the \citep{2009A&A...496..521C} fit within the context of the integrable Hamiltonian (\ref{Honedof}) is also shown as dashed curve. The considered system resides sufficiently deeply within the resonance that the trajectories never overlap with neighboring first-order resonances. Furthermore, as can be gleamed from the figure, the secular modulation of the system is weak. That is, the value of $\Upsilon$ never undergoes large variations. Consequently, the critical curve in the ($\Xi,\xi$) plane is never encountered and the pseudo-integral $\mathcal{J}$ is conserved along the trajectories. The dynamical portrait of the considered configuration is completely regular, (within the context of the second-order expansion of the disturbing function).}
\label{upsilonsectionCorr}
\end{figure}

Because the evolution timescales of the two degrees of freedom are very distinct, the distortions of the orbit in the ($\Xi,\xi$) plane that result from the modulation, preserve the adiabatic invariant, defined as \citep{1982amdc.proc..153H, 1984PriMM..48..197N}:
\begin{equation}
\mathcal{J} = \oint \Xi \ d \xi.
\end{equation}
Physically, the action $J$ represents an area occupied by a given orbit. The conservation of $J$ thus implies that any distortion of the orbit in the ($\Xi,\xi$) plane must be area-preserving. An important exception to this principle, intimately related to the onset of chaos, is that the conservation of $\mathcal{J}$ is broken when a trajectory encounters a critical curve. 

Consider an initially resonant orbit such those depicted by black lines in Figures (\ref{levelcurves2}). As long as the modulation of $\Upsilon$ is such that the inadmissible region remains far from the resonant orbit (e.g. taking panels A and B as the extremes of the modulation), the resonant region is not affected much. However, if we consider a stronger modulation (e.g. taking panels A and C as the extremes), it can be immediately seen that the area available for libration may shrink during a modulation cycle. In such a scenario, when the area of the separatrix becomes equal to the area occupied by the trajectory, the trajectory is forced to cross the separatrix, inevitably passing through the unstable (hyperbolic) equilibrium point. This marks the onset of chaotic motion. Following along the same lines of reasoning, one may deduce that if the separatrix disappears and reappears during a modulation cycle (e.g. taking panels C and D as the extremes), a considerable fraction (which depends on the modulation amplitude) of the trajectories may be understood to be chaotic.

With a handle on the role that the conservation (or lack thereof) of $\mathcal{J}$ plays, a nearly complete picture of the dynamics can be gleamed by sectioning the orbit in the ($\Xi,\xi$) plane and examining its evolution in the ($\Upsilon, \upsilon$) plane \citep{1985Icar...63..272W, 1990CeMDA..47...99H}. If the surface of section reveals a closed, regular orbit in the ($\Upsilon, \upsilon$) plane, it automatically implies that $\mathcal{J}$ is conserved along the evolutionary path and the separatrix was never encountered. Conversely, if the surface of section in ($\Upsilon, \upsilon$) plane reveals an area-filling manifold, conservation of $\mathcal{J}$ is broken as the orbit repeatedly encounters a separatrix in the ($\Xi,\xi$) plane \citep{1982amdc.proc..153H, 2002mcma.book.....M}. Indeed the situation is quite analogous to the well-studied problem of a modulated pendulum \citep{1991Nonli...4..615E, 1989PhyD...40..265B}.

\begin{figure}
\includegraphics[width=1\columnwidth]{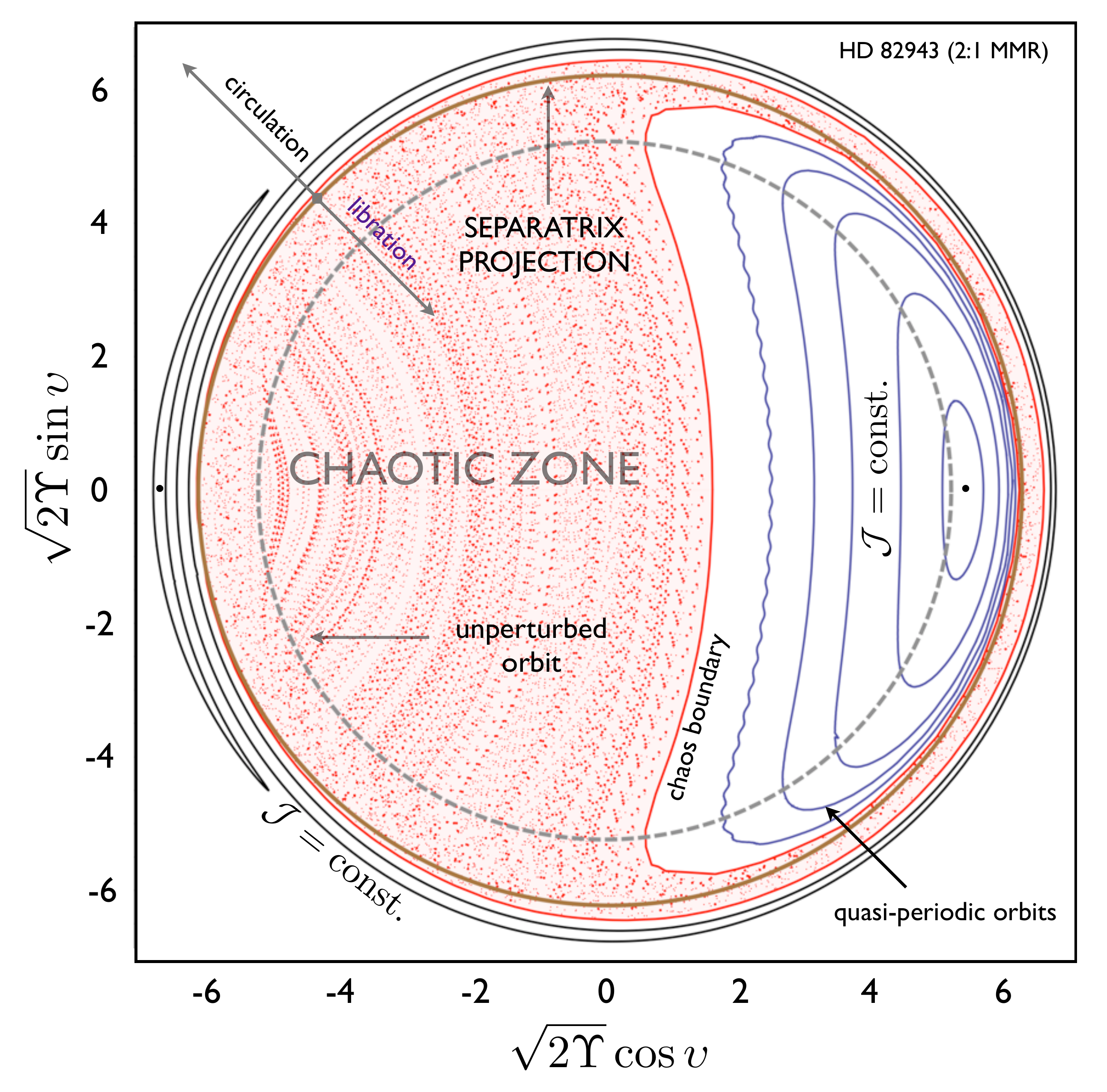}
\caption{A Poincare surface of section of the 2:1 mean motion resonance. The values of energy and angular momentum correspond to the dashed white line shown in the HD 82943 global dynamical map i.e. Figure (\ref{Poincare21}). Chaotic trajectories are shown with small red points while regular trajectories are depicted as purple and black curves. Conservation of the adiabatic invariant $\mathcal{J}$ is ensured by the separation of timescales along the shown regular trajectories. The orbits shown in purple are characterized by libration in the ($\Xi,\xi$) plane, while the black orbits imply circulation. Unsurprisingly, the ($\Upsilon,\upsilon$) phase-space occupied by orbits that entail libration in the ($\Xi,\xi$) plane are separated from those that entail circulation by the projection of the ($\Xi,\xi$) separatrix, shown as thick brown line. Additionally, as in Figure (\ref{upsilonsectionCorr}), the unperturbed orbit derived from the integrable Hamiltonian (\ref{Htwodof}) is shown as a dashed line.}
\label{upsilonchaos}
\end{figure}

 A surface of section of the \citep{2009A&A...496..521C} orbital solution is shown as a thick purple line in Figure (\ref{upsilonsectionCorr}). The apparent regularity of the observed motion raises the question if any ``relative" of the considered orbital fit, sharing the same values of $\mathcal{H}, \mathcal{K}$ and $\Omega$ can exhibit chaos. In order to address this, we surveyed the dynamical evolution of such orbits, characterized by different values of $\Upsilon$. A few examples of such orbits are plotted as thin purple lines on Figure (\ref{upsilonsectionCorr}). As can be gathered from the figure, the entire phase-space available to such orbits is occupied by regular trajectories. Evidently, any secular modulation of $\Upsilon$ permitted by the orbital energy and angular momentum corresponding to the \citep{2009A&A...496..521C} fit is not large enough to drive the orbit through a separatrix. This is not particularly surprising, since an examination of Figure (\ref{Poincare32}) explicitly shows that the energy level on which the \citep{2009A&A...496..521C} fit resides never approaches the vicinity of the separatrix. In other words, the \citep{2009A&A...496..521C} orbital solution is too deep within the resonance to exhibit chaotic motion. Note further that the circulation of $\upsilon$ seen in Figure (\ref{upsilonsectionCorr}) is fully consistent with libration of $\Delta \varpi$ seen in Figure (\ref{Correiacomp}). Indeed, if the asymmetry (that is controlled entirely by secular terms) of the orbit in the ($\upsilon,\Upsilon$) plane is not large, and the value of $\Upsilon$ (equivalently $\Phi_2$) remains sufficiently small (e.g. $\Phi_2$ exceeds the value corresponding to level A in Figure \ref{Poincare21}), the angle between the apsidal lines of the orbits remains in libration.

As already shown within the context of our discussion of resonance overlap, retaining the same starting level of $\Upsilon$ as that of the \citep{2009A&A...496..521C} fit and pushing the initial condition to an energy level that is close to the separatrix, will indeed result in highly irregular motion. However, the motion will not be irregular as a result of our sought-after effect, the secular modulation. Consequently, in order to demonstrate the onset of chaos due to secular interactions more coherently, let us relocate our discussion to the 2:1 resonance and choose a starting value of $\Upsilon$ and an energy level such that the unperturbed solution lies close to the 2:1 separatrix yet far enough away from the 3:2 separatrix for the perturbations from the neighboring resonance to rapidly average out. 

The initial condition we shall consider lies on a contour of $\Upsilon$ directly above the one labeled B in Figure (\ref{Poincare21}) and on the energy level that intersects the contour immediately inside the separatrix. For convenience, the unperturbed version of the starting state in question is labeled by a dashed white line in Figure (\ref{Poincare21}). Naturally, as can be inferred from the figure, the unperturbed solution is characterized by large-amplitude resonant libration in the $(\Xi,\xi)$ plane. 

Accounting for the secular terms, the evolution of this initial condition exhibits large-scale chaos. The extensive chaotic sea occupied by the solution is shown with opaque red points in the Poincare surface of section (\ref{upsilonchaos}). However, the phase-space portrait is not entirely occupied by irregular trajectories. A survey of initial conditions permitted by the values of energy and angular momentum reveals the existence of quasi-periodic solutions as well, depicted as purple and black curves.  $\mathcal{J}$ is conserved at all times along these curves. 

The dynamics is characterized by resonant libration in the $(\Xi,\xi)$ plane within the regular region on the inside of the chaotic zone (corresponding to purple curves in Figure \ref{upsilonchaos}) and by circulation in the $(\Xi,\xi)$ plane on the outside of the chaotic zone (corresponding to black curves). It should be noted that the example considered here was specifically chosen to reside in close proximity to the separatrix. For an arbitrary choice of initial conditions, even if the chaotic zone is permissible by the conservation of angular momentum and energy, it would likely occupy a considerably smaller fraction of phase-space. 

The boundary between circulation and libration is denoted by the thick brown circle shown in Figure \ref{upsilonchaos}). In other words, the thick brown circle is a projection of the separatrix in the $(\Xi,\xi)$ plane onto the $(\Upsilon,\upsilon)$ plane. The attribution of the origin of chaos to secular modulation is exemplified by the fact that the projected separatrix hugs the boundary of the chaotic zone on the inside as well as the outside. Thus, any secular trajectory that crosses the projected separatrix is driven to irregularity. 

It is worth noting that although any orbit that starts out within the chaotic sea will be irregular by definition, unlike the case of mean motion resonance overlap considered above, between encounters with the separatrix, the evolution will be characterized by conservation of $\mathcal{J}$ and will therefore be temporarily regular. Thus, a clear difference between the two chaotic regimes can be established. Chaos that arises from secular modulation is described by slow diffusion that takes place on a secular timescale, while diffusion that arrises from the overlap of mean motion resonance is fast, characterized by the resonant timescale.

\section{Divergent Resonant Encounters}

An interesting and useful application of the theory formulated above is the treatment of divergent encounters of planets with mean motion resonances. As briefly described in the introduction, capture into resonance requires convergent migration \citep{1984CeMec..32..127B, 1991pscn.proc..193H}. In contrast, resonant encounters that stem from divergent migration can never lead to capture and instead always yield impulsive excitation of the orbits. Here, we wish to consider the latter scenario and address the translation of post encounter dynamics onto the secular domain. 

While studying divergent encounters with the 2:1 mean motion resonance by Jupiter and Saturn within the context of the Nice model, \citet{2009A&A...507.1041M} identified that the planets always come out of the resonance locked in an apsidally anti-aligned state. Furthermore,  the apsidal alignment persists indefinitely, unless it is broken by a close encounter with a transiently unstable ice-giant. Although \citet{2009A&A...507.1041M} attributed the origin of the apsidal lock to a fortunious mass-ratio between Jupiter and Saturn, here we assert that this result is largely independent of the planetary masses.

\begin{figure}
\includegraphics[width=1\columnwidth]{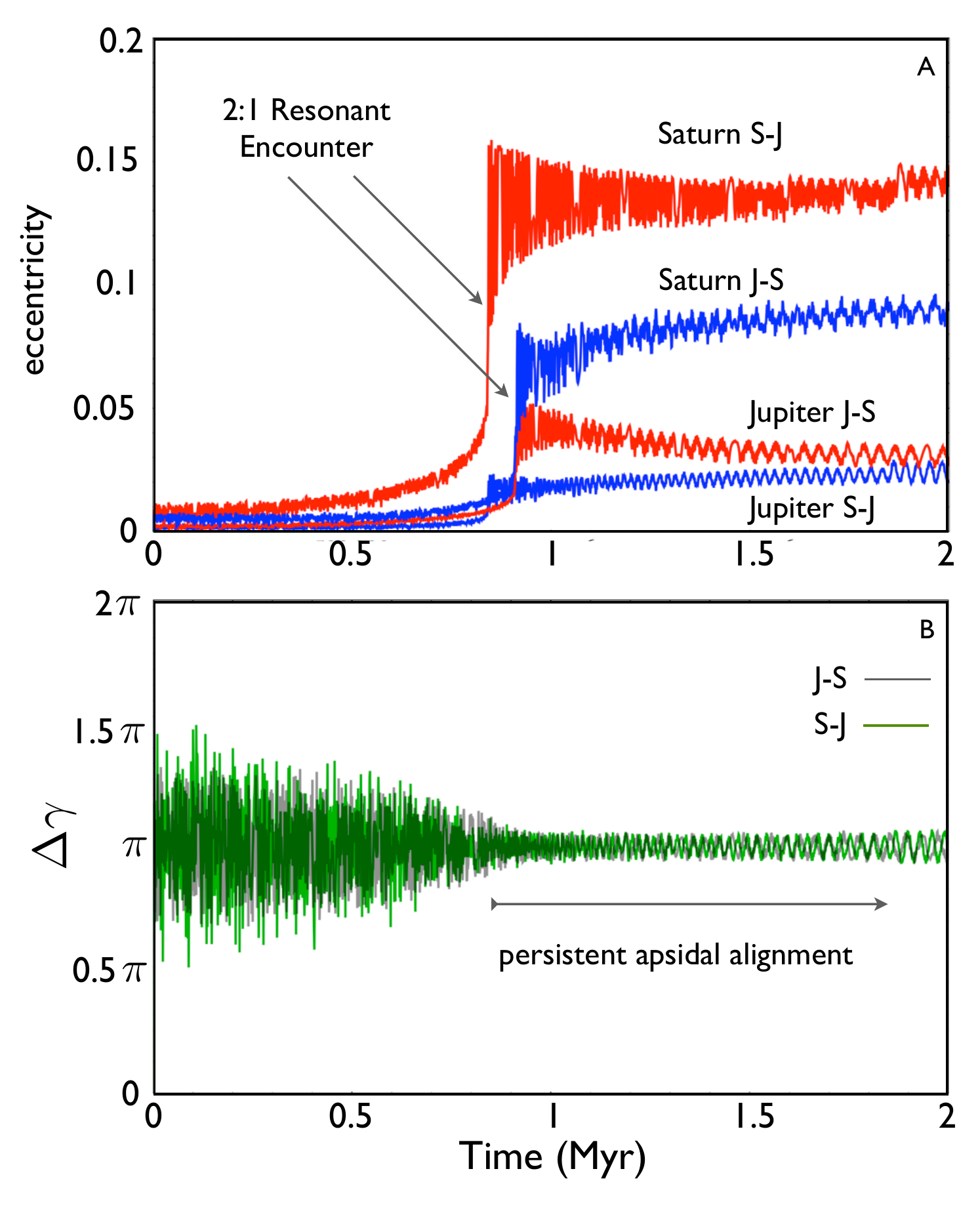}
\caption{Impulsive excitation of the orbits by an encounter with the 2:1 mean motion resonance. This figure depicts the results of two numerical experiments where a Jupiter-mass planet and a Saturn-mass planet encounter a 2:1 mean motion resonance. Panel A shows the evolution of the eccentricities while panel B shows the evolution of the difference in the perihelia. In one experiment (labeled JS), Jupiter is placed on the inner orbit. In the other numerical experiment (labeled SJ), Saturn resides closer to the Sun. Although the extent of eccentricity excitation differs between the two runs considerably, the post-encounter apsidal alignment is clearly evident in both cases. }
\label{dvarpi}
\end{figure}

Panels A and B in Figure (\ref{dvarpi}) show the time evolution of the eccentricities and the difference in longitudes of perihelia where the 2:1 mean motion resonance is encountered by Jupiter and Saturn as well as a planetary pair with reversed masses i.e. Jupiter residing further from the sun. The solutions are obtained from numerical experiments where divergent migration was implemented via a fictitious force. Specifically, the simulations were performed using the Symba N-body integration software package \citep{1998AJ....116.2067D}, modified such that in isolation, the outer orbit drifts outwards and the inner orbit drifts inwards with the migration rate decaying as $\propto \exp{t/\tau}$, choosing $\tau = 1$Myr \citep{2009A&A...507.1041M}. As can be assessed from the figure, the values of the eccentricities acquired by the planets during the resonance passage depend on the planetary mass ratio. Yet the encounter drives the planets to an apsidally anti-aligned state in both cases. Changing the mass ratio to unity does not affect the results.

\subsection{Pre-Encounter Evolution}

This behavior can be readily explained in the context of the model developed here. Let us begin by first discussing the pre-encounter initial conditions. Because orbital migration is usually driven by time-irreversible (effectively dissipative) processes (e.g. interactions with the protoplanetary nebula, tidal interactions, planetesimal scattering), it is natural to assume that planets migrate on circular orbits. As a result, recalling the definitions of the variables (\ref{amdactionangle}), we shall set the pre-encounter actions (where planets reside far away from resonance) to $\Xi^{\rm{pre}} = \Upsilon^{\rm{pre}} = 0$. 

Next, consider the migration rate. Numerical simulations \citep{2005Natur.435..459T, 2007A&A...461.1173C, 2008Icar..193..267Z} suggest that in most cases of interest, the rate of orbital migration is slow compared to the secular interaction timescale, closely related to the evolution timescale of the acton $\Upsilon$. Taking the assumption of slow migration as a guiding principle, we are tempted to define a second adiabatic invariant related to secular motion. However, prior to doing so, we must first examine if the adiabatic approach is viable despite near-null eccentricities, which we showed in the last section can lead to the appearance of secondary resonances.

At first glance, adiabatic invariance seems impossible because the criterion given by equations (\ref{secres1}) and (\ref{dpsi2dt}) clearly indicates that the system should be close to the 1:1 secondary resonance (see also Figure \ref{secondaryres}). However, as already mentioned in the previous section, in the limit of a vanishingly low value of $\Upsilon$, the criterion for secondary resonances must be reevaluated in light of the possibility of a librating rather than circulating $\upsilon$. The procedure we follow is essentially identical to that outlined in section (3.2.1), but in order to appropriately capture the dynamics, we must work with a two degree of freedom Hamiltonian. Specifically, we shall consider a simplified version of Hamiltonian (\ref{Htwodof}), where the secular terms are dropped with the exception of the harmonic\footnote{Because the eccentricities are taken to be very low, retention of secular terms that are linear in the actions does not change the results in any meaningful way, while making the already formidable algebra even more opaque.}. Assembling the relevant constants to the secular part of the Hamiltonian into a single constant $\mathcal{C}_{\rm{s}}$ (see equation \ref{Htwodof}), we have:
\begin{eqnarray}
\mathcal{H} &=& \hat{\delta} (\Omega + \Xi) - (\Omega + \Xi) ^2 - \sqrt{1- \Upsilon/\Xi} \sqrt{2 \Xi} \cos(\xi) \nonumber \\
&-& \mathcal{C}_{\rm{s}} \sqrt{\Xi/\Upsilon -1 } \Upsilon \cos (\upsilon) . 
\end{eqnarray} 

Upon expanding the Hamiltonian in Taylor series to second order around nominal resonance in both degrees of freedom (that is, ($\Xi,\xi$) is expanded around ($[\Xi],0$) and ($\Upsilon,\upsilon$) is expanded around ($[\Upsilon],0$)), we obtain the following expression:
\begin{eqnarray}
\label{thingy}
\mathcal{H} &=& \frac{\sqrt{2}}{2} \sqrt{[\Xi] - [\Upsilon]} \bar{\xi}^2 + \frac{1}{2} \bigg( \frac{1}{2 \sqrt{2} ([\Xi] -  [\Upsilon])^{3/2}} \nonumber \\
&-& \frac{\mathcal{C}_{\rm{s}} \sqrt{[\Upsilon]}}{4 ([\Xi] -  [\Upsilon])^{3/2}} -2 \bigg) \bar{\Xi}^2 - \frac{\mathcal{C}_{\rm{s}} \sqrt{[\Upsilon]}  }{2} \sqrt{[\Xi] - [\Upsilon]} \bar{\upsilon}^2 \nonumber \\
&-& \frac{1}{2} \bigg(\frac{\mathcal{C}_{\rm{s}} [\Xi]^2}{4 [\Upsilon]^{3/2} ([\Xi]-[\Upsilon])^{3/2}} - \frac{1}{2 \sqrt{2} ([\Xi]-[\Upsilon])^{3/2} } \bigg) \bar{\Upsilon}^2  . 
\end{eqnarray}
where the barred variables are defined as the deviations away from equilibrium (see equation (\ref{equilibvars}) for an analogous definition). As before, the nominal actions are given by setting the linear terms in the above Hamiltonian to zero:
\begin{eqnarray}
\label{lineqs}
\hat{\delta} - 2 [\Xi] - 2 \Omega + \frac{\mathcal{C}_{\rm{s}} \sqrt{[\Upsilon]}}{\sqrt{[\Xi]-[\Upsilon]}} - \frac{1}{\sqrt{2} \sqrt{[\Xi]-[\Upsilon]}}=0, \nonumber \\
\frac{1}{\sqrt{2} \sqrt{[\Xi]-[\Upsilon]}} - \frac{\mathcal{C}_{\rm{s}} \sqrt{[\Upsilon]}}{\sqrt{[\Xi]-[\Upsilon]}} + \frac{\mathcal{C}_{\rm{s}}[\Xi]}{2 \sqrt{[\Upsilon]} \sqrt{[\Xi]-[\Upsilon]}}=0. 
\end{eqnarray}

We are now in a position to convert the Hamiltonian (\ref{thingy}) into the form of two decoupled harmonic oscillators. However, before doing so let us examine equations (\ref{lineqs}) in greater detail. Adding the two equations together, we can obtain an expression for $[\Upsilon]/[\Xi]$:
\begin{eqnarray}
\frac{[\Upsilon]}{[\Xi]} = \frac{\mathcal{C}_{\rm{s}}^2}{\mathcal{C}_{\rm{s}}^2 + 4 (\hat{\delta}-2([\Xi]+\Omega))^2} \simeq  [\Xi] \frac{\mathcal{C}_{\rm{s}}^2}{2}.
\end{eqnarray}
The latter simplification utilizes the fact that $\mathcal{C}_{\rm{s}} \ll \hat{\delta} - 2 ([\Xi] + \Omega)$ since the former arises from a higher order perturbation. Note that this expression implies that $[\Upsilon]/[\Xi]$ is a small parameter. This relationship between $[\Upsilon]$ and $[\Xi]$ will prove useful in obtaining simplified expressions for the libration frequencies below.

Employing a change of variables of the same type as (\ref{sqrttrans}) with coefficients from equation (\ref{thingy}), we transform the Hamiltonian into the desired form:
\begin{eqnarray}
\label{doubleho}
\mathcal{H} = \frac{\varphi_{\xi}}{2}(\tilde{\Xi}^2 + \tilde{\xi}^2) + \frac{\varphi_{\upsilon}}{2}(\tilde{\Upsilon}^2 + \tilde{\upsilon}^2) . 
\end{eqnarray}

The explicit expressions for the libration frequencies $\varphi_{\xi}$ and $\varphi_{\upsilon}$ can be made simpler by expanding them to leading order in $[\Upsilon]/[\Xi]$, which we showed above to be a small parameter:
\begin{eqnarray}
\varphi_{\xi} = \sqrt{\frac{\sqrt{2}-8([\Xi]-[\Upsilon])^{3/2}- \mathcal{C}_{\rm{s}} \sqrt{[\Upsilon]}}{2 \sqrt{2} ([\Xi]-[\Upsilon])}} \simeq \sqrt{\frac{1}{2[\Psi]_1}}, \nonumber \\
\varphi_{\upsilon} = \frac{1}{2} \sqrt{\frac{\mathcal{C}_{\rm{s}} (\sqrt{2}[\Upsilon]^{3/2}-[\Xi]^2 \mathcal{C}_{\rm{s}})}{[\Upsilon]([\Upsilon]-[\Xi])}} \simeq \frac{\mathcal{C}_{\rm{s}} \sqrt{[\Xi]}}{2 \sqrt{[\Upsilon]}} \simeq \sqrt{\frac{1}{2[\Psi]_1}}.
\end{eqnarray}
Evidently, the two angles, $\xi$ and $\upsilon$ evolve on similar timescales. As in the previous section, we can take advantage of this similitude to easily identify an adiabatic invariant.

Let us implicitly define two sets of action angle coordinates
\begin{eqnarray}
\tilde{\Xi} = \sqrt{ 2 \tilde{\mathcal{X}}} \cos{\tilde{x}} \ \ \ \ \ \ \tilde{\xi}= \sqrt{ 2 \tilde{\mathcal{X}}} \sin{\tilde{x}} , \nonumber \\
\tilde{\Upsilon} = \sqrt{ 2 \tilde{\mathcal{Y}}} \cos{\tilde{y}} \ \ \ \ \ \  \tilde{\upsilon}= \sqrt{ 2 \tilde{\mathcal{Y}}} \sin{\tilde{y}}.
\end{eqnarray}
In these variables, the Hamiltonian (\ref{doubleho}) reads:
\begin{eqnarray}
\mathcal{H} = \varphi_{\xi} \tilde{\mathcal{X}} + \varphi_{\upsilon} \tilde{\mathcal{Y}}. 
\end{eqnarray}
Applying a contact transformation originating from the generating function 
\begin{eqnarray}
F_{2} = \tilde{x} \tilde{\mathcal{W}} + (\tilde{y}-\tilde{x}) \tilde{\mathcal{Z}},
\end{eqnarray}
we obtain the action-angle variables:
\begin{eqnarray}
\tilde{\mathcal{W}} = \tilde{\mathcal{X}}+\tilde{\mathcal{Y}}, \ \ \ \ \ \ \ \ \tilde{w} = \tilde{x}, \nonumber \\
\tilde{\mathcal{X}} =  \tilde{\mathcal{Y}}, \ \ \ \ \ \ \ \ \ \ \ \ \tilde{z} = \tilde{y} - \tilde{x}.
\end{eqnarray}

The Hamiltonian is now explicitly adiabatic:
\begin{eqnarray}
\mathcal{H} = \varphi_{\xi} \tilde{\mathcal{W}} + (\varphi_{\upsilon} - \varphi_{\xi}) \tilde{\mathcal{Z}}. 
\end{eqnarray}
Indeed, the evolution of the angle $\tilde{z}$ is much slower than that of the angle $\tilde{w}$. This allows us to reintroduce the first adiabatic invariant
\begin{equation}
\mathcal{J} = \oint \tilde{\mathcal{W}} \ d \tilde{w}.
\end{equation}
Moreover, assuming that migration occurs more slowly than the evolution of $\tilde{z}$, we can define a second adiabatic invariant\footnote{The distinction between the first and the second adiabatic invariants $\mathcal{J}$ and $\mathcal{I}$ parallels the analogous definitions employed in plasma physics \citep{2006fpp..book.....B}. 
That is, the second invariant corresponds to a much longer timescale than that of the first, and is therefore broken more easily.}
\begin{equation}
\mathcal{I} = \oint \tilde{\mathcal{Z}} \ d \tilde{z}.
\end{equation}

In essence, the system we are concerned with here is subject to the double-adiabatic condition. Namely, $\mathcal{I}$ is conserved by construction because the migration rate is taken to be sufficiently slow and $\mathcal{J}$ is conserved because the two degrees of freedom are well-separated. Although the secular phase space portrait depicted by the Hamiltonian (\ref{Htwodof}) contains no critical curves, conservation of both, $\mathcal{J}$ and $\mathcal{I}$ is broken when a separatrix is encountered in ($\Xi,\xi$) space. Consequently, the double-adiabatic condition applies before and after, but not during the resonant encounter. However, because the impulsive excitation of the orbits occurs on a resonant timescale, $\Upsilon$ (and equivalently, $\Psi_2$) itself is conserved across the encounter. 

At this point, we have enough information to show that after the resonant encounter, the orbits must be anti-aligned. Let us begin by reasoning through the calculation of the impulsive orbital excitation.

Throughout the evolution prior to the encounter, $\mathcal{J} = \mathcal{I} = 0$. Note that this condition does not imply circular orbits. Instead, it implies that the system resides on a global fixed point, nearest to the origin.
 
As exact resonance is approached, the location of the fixed point on the ($\Xi,\xi$) plane moves to the right (i.e. acquires a finite value of $\Xi$ while remaining at $\xi = 0$). In other words, the approach to exact resonance can be viewed as sequential evolution through panels A, B, C and D of Figure (\ref{levelcurves}), where the solution resides on the black dot in the center of the inner circulation zone. As the inner circulation zone contracts, the stable fixed point and the unstable fixed move closer together in phase space. 

\subsection{Post-Encounter Evolution}

\begin{figure}
\includegraphics[width=1\columnwidth]{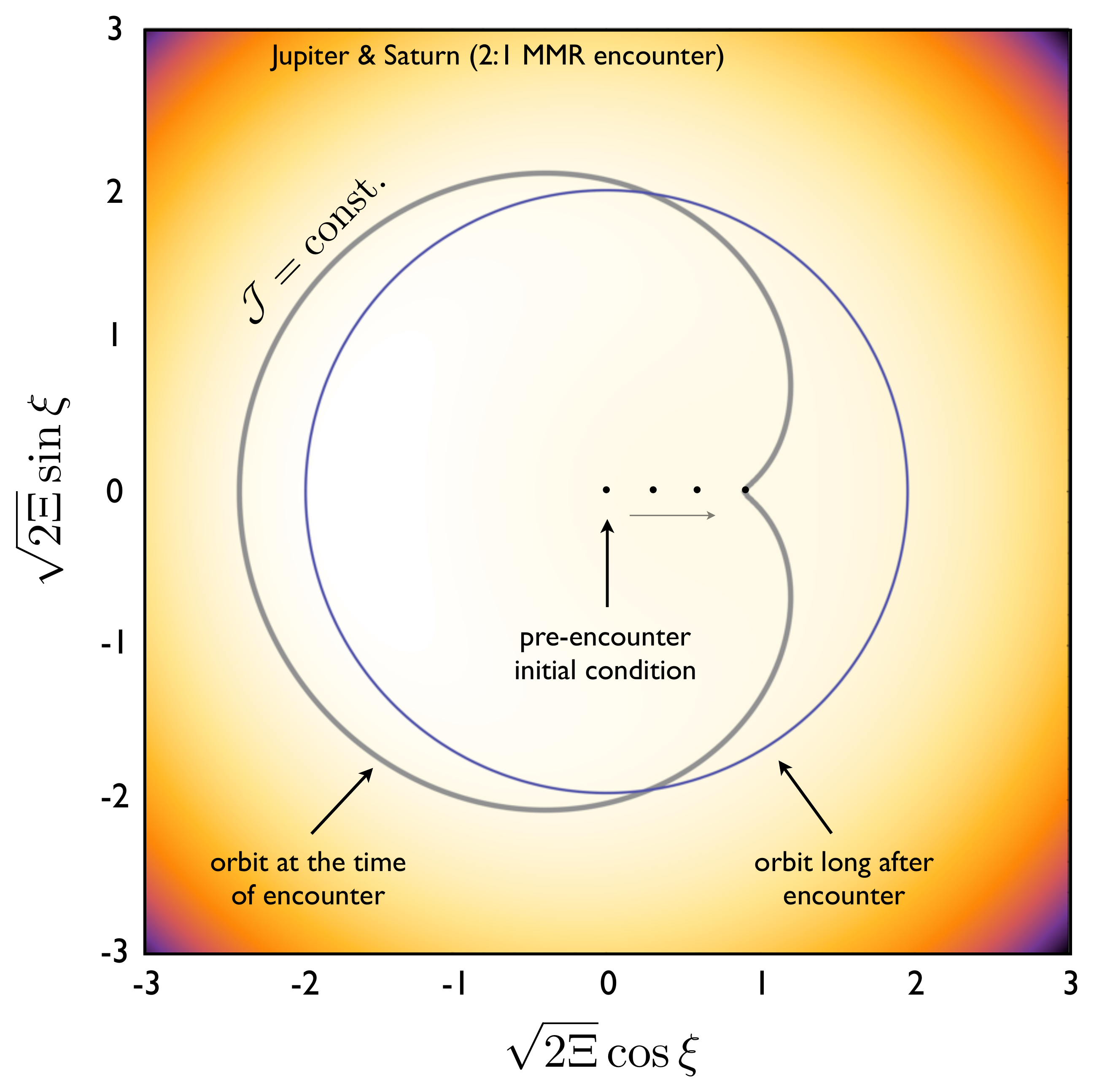}
\caption{Phase-space representation of the divergent resonant encounter. Prior to the encounter, the system resides on the stable equilibrium point in the vicinity of the origin. At the time of the encounter, the stable equilibrium point and the unstable equilibrium point (on which the separatrix resides) join and the system obtains a circulational trajectory, related to the separatrix. As the system marches further away from resonance, the circulating trajectory asymptotically approaches a circle, while conserving the encapsulated area, $\mathcal{J}$.}
\label{JSencounter}
\end{figure}

The impulsive excitation occurs when the stable fixed point on which the dynamics resides and the unstable fixed point at the crest of the separatrix join. As long as the resonant encounter takes place at low eccentricities, the phase-space portrait of the system can be visualized, neglecting the second order secular contribution. In the purely resonant framework, this occurs when two of the roots to the cubic equilibrium equation, derived from the Hamiltonian (\ref{Honedof})
\begin{equation}
1+ \rho^3 + 2 \Omega \rho = \hat{\delta} \rho,
\end{equation}
where $\rho = \sqrt{2 \Xi}$, are identical \citep{1999ssd..book.....M}. In fact, the bifurcation of the fixed point can be used to calculate the exact semi-major axis ratio at which the encounter occurs. At this point, conservation of $\mathcal{J}$ is momentarily broken and the system obtains an orbit defined by the separatrix in the ($\Xi,\xi$) plane, shown as a gray curve in Figure (\ref{JSencounter}). 

Because the resonant encounter occurs ``instantaneously" with respect to the migration timescale, the orbital angular momentum must be conserved across the jump. Consequently, the acquisition of angular momentum deficit (related to $\Xi$) is accompanied with a small jump in the semi-major axis ratio that converts the separatrix into a similarly-shaped regular circulating orbit. The circulating, rather than librating nature of the new orbit is ensured because during divergent migration, the phase-space area occupied by resonant trajectories shrinks, preventing capture \citep{1982amdc.proc..153H, 1986sate.conf..159P}. Strictly speaking, this means that the dynamics no longer resides on a fixed point in the ($\Upsilon,\upsilon$) plane because the newly acquired angular momentum deficit changes the dynamical portrait. Indeed, the new trajectory in the ($\Upsilon,\upsilon$) plane envelopes the new coordinates of the fixed point and passes through the pre-encounter equilibrium location. However, it can be argued with some level of rigor that change in the fixed point's location will be small and by extension, so will the radius of the post-encounter orbit in the ($\Upsilon,\upsilon$) plane.

First, note that neglecting second order terms, the ($\Upsilon,\upsilon$) fixed point always resides at the origin because the Hamiltonian (\ref{Honedof}) is independent of $\upsilon$ (equivalently, $\psi_2$). This line of reasoning is a useful starting point but is an oversimplification as it only implies trivial secular dynamics embedded in the transformation (\ref{suicide}). In reality (as can be seen in Figure \ref{upsilonsectionCorr}), the ($\Upsilon,\upsilon$) fixed point resides somewhat off-center. In particular, prior to the encounter, the fixed point in the ($\Upsilon,\upsilon$) plane is obtained from equations (\ref{lineqs}).

After the encounter, the ($\Upsilon,\upsilon$) equilibrium point can be calculated in a similar way, however, taking into account the fact that ($\Xi,\xi$) no longer resides at an equilibrium point. Consider a modified version of equation (\ref{lineqs}):
\begin{equation}
\label{timeav}
\frac{1}{T}\int_0^T \left[ \frac{\cos(\xi)}{\sqrt{2}\sqrt{\Xi-\Upsilon}} - \frac{C_s\sqrt{\Upsilon}}{\sqrt{\Xi-\Upsilon}} + \frac{C_s\Xi}{(2\sqrt{\Upsilon}\sqrt{\Xi-\Upsilon})} \right]= 0,
\end{equation}
where $T$ is the period required to complete a single orbit in the ($\Xi,\xi$) plane. Note that the above expression simplifies to equation (\ref{lineqs}b) in the limit where $\Xi=[\Xi]$ and $\xi=0$ for all $t$.

As already stated above, immediately after resonance crossing the ($\Xi,\xi$) trajectory begins circulation (see Figure \ref{JSencounter}). However, during a single circulation cycle, the ($\Xi,\xi$) trajectory spends most of its period in close proximity to the ($[\Xi],0$) fixed point, because that is where the time derivative of $\xi$ is minimal. Thus, the solution of equation (\ref{timeav}) in $\Upsilon$ will be close to that of equation (\ref{lineqs}). In other words, the equilibrium point in ($\Upsilon,\upsilon$) will not move considerably. Consequently, the dynamics of ($\Upsilon,\upsilon$), which was on the stable equilibrium point before the resonance crossing, will describe a cycle around the new equilibrium point after the crossing. The corresponding radius of the orbit will equal to the displacement suffered by the equilibrium point itself, which is small.

For all subsequent evolution, as the planets migrate away from resonance, the conservation of both adiabatic invariants is once again in effect. Consequently, because of the conservation of the first adiabatic invariant $\mathcal{J}$, the orbit on the ($\Xi,\xi$) plane asymptotically approaches a circle centered on the origin, whose area is given by:
\begin{equation}
2 \pi \Xi = \mathcal{J}_{\rm{separatrix}} .
\end{equation}

Moreover, on the ($\Upsilon,\upsilon$) plane, the small radius of the cycle around the equilibrium point will be maintained, thanks to the conservation of the second adiabatic invariant, $\mathcal{I}$. As long as this equilibrium point  remains close to the origin, $\upsilon$ may circulate, but the smallness of $\Upsilon$ ensures that $\Delta\varpi$ librates around $\pi$ (see $\Upsilon = \Psi_2 \simeq 0$ contours in Figures \ref{Poincare21} and \ref{Poincare32}). In principle, as the planets move away from resonance, the ($\Upsilon,\upsilon$) equilibrium point can move away from the origin. In this case, the small radius of the orbit around the equilibrium point on the $(\Upsilon,\upsilon$) plane implies that $\upsilon$ librates. This, again, ensures the libration of $\Delta\varpi$ around $\pi$.

As a final point, it is important to comment on the results of an additional numerical experiment reported by \citet{2009A&A...507.1041M}. In particular, \citet{2009A&A...507.1041M} showed that if the masses of both, Jupiter and Saturn are reduced by a factor of 100, the post-encounter apsidal alignment among the orbits no longer holds. This phenomenon (although apparently contradictory to the statements made above), can also be understood within the context of our model. Recall that our formulation of resonant encounters specifically assumed the double adiabatic condition. In the low-mass experiment considered by \citet{2009A&A...507.1041M}, the conservation of $\mathcal{I}$ is broken because the migration timescale is taken to be faster than the longest interaction timescale of the planets. Consequently, we can expect that there exists a tentative cut-off in mass below which apsidal alignment cannot endure. The characteristic value of such a cutoff however is dependent on the migration process in question and will therefore vary among differing astrophysical settings. 

\section{Conclusion}

In this paper, we have set out to construct a simple geometrical representation of the global dynamics of the unrestricted, first order resonant three-body problem. As the primary purpose of the paper is the delineation of a comprehensive dynamical picture, we have opted to work within the context of analytically tractable, but approximate perturbation theory.

Although first-order resonant motion can be apparently complex, here, greatly aided by the pioneering works of \citet{1984CeMec..32..307S} as well as \citet{1986CeMec..38..335H} and \citet{1986CeMec..38..175W}, we have shown that the essential features of the dynamics is captured within the context of a simple integrable Hamiltonian. The Hamiltonian in question is qualitatively similar to that of a pendulum and more precisely, is related to the second fundamental model for resonance \citep{1983CeMec..30..197H}. This highlights a certain kinship between the unrestricted and the restricted three-body problems, as the second fundamental model for resonance has also been applied extensively to the study of the latter.  

Quantitatively, the formulated theory is only accurate at low eccentricities. Nevertheless, it still provides the much-needed qualitative insight relevant to a broad range of orbital architectures. Indeed, at an age when N-body integration software is freely available \citep{1998AJ....116.2067D, 1999MNRAS.304..793C} and computational resources required for problems such as these are abundant, the qualitative understanding that emerges from the theory is of greater importance than the particularities of its direct application. Consequently, the utility of the developed theory is best envisioned as a theoretical supplement to (rather than a replacement of) numerical N-body simulations.

Utilizing the various constants of motion that arise within the context of the integrable theory, we have constructed a geometrical characterization of the resonant motion. Indeed, global maps of the dynamics, such as those presented in Figures (\ref{Poincare21}) and (\ref{Poincare32}) provide a visual aid that allows one to instantly assess important features of any particular resonant solution such as the proximity of the system to a separatrix or conversely the depth within the resonance at which a given orbital fit resides. Although the global maps (\ref{Poincare21}) and (\ref{Poincare32}) are restricted by the fact that they portray surfaces of section, combined with corresponding phase-space portraits, such as those presented in Figures (\ref{levelcurves}) and (\ref{levelcurves2}), a more comprehensive understanding of the dynamics can be obtained.  

The applicability of the integrable theory is unavoidably limited. An important, well-known feature of resonant dynamics is its capacity for chaotic motion. Because the nature of the integrable model is inherently regular, in isolation, it is essentially of no use in the chaotic domain.

In this work, we emphasized two distinct modes of the onset of chaos. Namely, we considered the rapid irregularity that arises from the  overlap of mean motion resonances as well as slow chaos that arises as a result of the secular modulation of the orbit through the separatrix. The first mode dominates in the region of parameter space where neighboring resonant separaticies reside in some proximity to each other. In direct analogy with the restricted problem, for ($k:k-1$) resonances, the region of parameter space occupied by this effect grows with increasing $k$. Conversely, chaotic diffusion near seaparatricies that are isolated from neghboring mean motion resonances is dominated by secular modulation of the resonant dynamics. 

It is important to recall that beyond the integrable approximation, we only accounted for a limited number of second order secular terms. Obviously, even after averaging out short-period terms, the dynamics encapsulated into the residual disturbing function is much richer than the simple model utilized here. This implies that the description of the onset of chaos is far from exhaustive. That said, the method outlined in this paper, namelyintroducing an integrable Hamiltonian by freezing the secular degree of freedom and then studying its evolution in the adiabatic regime, is valid for arbitrary eccentricities and inclinations. Consequently, the largely qualitative account of the onset of chaos presented here should be viewed as a guide to a general methodology rather than a particular model with extended applicability.

As an application of the simple theory formulated in this work, we addressed divergent resonant encounters between massive planets. Particularly, we showed that the natural outcome of adiabatic resonant encounters is an apsidally anti-aligned orbital state. Interestingly, this result is largely independent of the planetary masses. Moreover, the preservation of the second adiabatic invariant (related to secular dynamics) ensures that small-amplitude libration around the anti-aligned fixed point persists far away from the resonance.

As a consequence of this result, it is tempting to interpret small-amplitude anti-aligned libration of nonresonant planets as a signature of past resonant encounters as well as the associated migration. Indeed, such an interpretation holds great value as an instrument for disentangling the dynamical histories of planetary systems. However, care must be taken when drawing any such conclusion because eccentricity damping (such as that resulting from the dissipative processes that drive divergent orbital migration in the first place) in the secular domain may lead to anti-aligned orbits independently \citep{2002ApJ...564.1024W, 2007MNRAS.382.1768M, 2011ApJ...730...95B}. Furthermore, it is important to note that lack of co-precessing anti-aligned obits in a given system should not be viewed as evidence for lack of past resonant encounters, since resonant encounters in densely populated planetary systems, can lead to orbital instabilities that  act to chaotically erase fossilized remnants of past evolution. The lack of apsidal alignment between Jupiter and Saturn suggests that the solar system is in fact, such an example \citep{2009A&A...507.1041M, 2010ApJ...716.1323B, 2012AJ....144..117N}.

Although we have solely addressed divergent resonant encounters here, the same model can also be applied to convergent resonant encounters. As discussed above, the outcomes of convergent encounters include both, capture into resonance as well as capture-free orbital excitations \citep{1991pscn.proc..193H, 2002ApJ...567..596L}. While the latter scenario is qualitatively similar to the example considered here, in case of successful capture, post-encounter evolution can depend strongly on factors such as the orbital migration and eccentricity dissipation rates as well as the strength of external stochastic perturbations \citep{2008ApJ...683.1117A, 2010A&A...510A...4R}. Similar factors contribute to the determination of whether capture can occur in the first place \citep{1999ssd..book.....M}. 

Consequently, astrophysically relevant analysis of convergent resonant encounters within the framework of the model discussed here requires extensive, numerical validation. Owing to the significant associated computational cost of such a project, addressing this issue is far beyond the scope of the current study. However, our investigation aimed at quantifying the various regimes of convergent resonant encounters is already underway and will be published in a subsequent follow up study. 
\\
\\
\\

%\begin{acknowledgments} 
We wish to thank Jake Ketchum and Matt Holman for carefully reviewing the manuscript and providing helpful suggestions. We wish to thank Christian Beauge for a very thorough and insightful referee report that greatly enhanced the quality of this manuscript. K.B. acknowledges the generous support from the ITC Prize Postdoctoral Fellowship at the Institute for Theory and Computation, Harvard-Smithsonian Center for Astrophysics.  
%\end{acknowledgments}

{}


\begin{thebibliography}{}

\bibitem[Adams et al.(2008)]{2008ApJ...683.1117A} Adams, F.~C., Laughlin, 
G., Bloch, A.~M.\ 2008.\ Turbulence Implies that Mean Motion Resonances are 
Rare.\ The Astrophysical Journal 683, 1117-1128. 

\bibitem[Armitage(2010)]{2010apf..book.....A} Armitage, P.~J.\ 2010. Astrophysics of Planet Formation, by 
Philip J.~Armitage, pp.~294.~ISBN 978-0-521-88745-8 (hardback).~Cambridge, 
UK: Cambridge University Press, 2010.\ .

\bibitem[Batygin and Brown(2010)]{2010ApJ...716.1323B} Batygin, K., Brown, 
M.~E.\ 2010.\ Early Dynamical Evolution of the Solar System: Pinning Down 
the Initial Conditions of the Nice Model.\ The Astrophysical Journal 716, 
1323-1331. 

\bibitem[Batygin and Laughlin(2011)]{2011ApJ...730...95B} Batygin, K., 
Laughlin, G.\ 2011.\ Resolving the sin(I) Degeneracy in Low-mass 
Multi-planet Systems.\ The Astrophysical Journal 730, 95. 

\bibitem[Batygin et al.(2011)]{2011ApJ...738...13B} Batygin, K., Brown, 
M.~E., Fraser, W.~C.\ 2011.\ Retention of a Primordial Cold Classical 
Kuiper Belt in an Instability-Driven Model of Solar System Formation.\ The 
Astrophysical Journal 738, 13. 

\bibitem[Batygin and Morbidelli(2012)]{2012arXiv1204.2791B} Batygin, K., 
Morbidelli, A.\ 2012.\ Dissipative Divergence of Resonant Orbits.\ ArXiv 
e-prints arXiv:1204.2791. 

\bibitem[Beauge(1994)]{1994CeMDA..60..225B} Beauge, C.\ 1994.\ Asymmetric 
liberations in exterior resonances.\ Celestial Mechanics and Dynamical 
Astronomy 60, 225-248.

\bibitem[Bellan(2006)]{2006fpp..book.....B} Bellan, P.~M.\ 2006.\ Fundamentals of Plasma Physics, by Paul 
M.~Bellan, pp.~.~ISBN 0521821169.~Cambridge, UK: Cambridge University 
Press,  2006.\ . 

\bibitem[Bruhwiler and Cary(1989)]{1989PhyD...40..265B} Bruhwiler, D.~L., 
Cary, J.~R.\ 1989.\ Diffusion of particles in a slowly modulated wave.\ 
Physica D Nonlinear Phenomena 40, 265-282. 

\bibitem[Bitsch and 
Kley(2011)]{2011A&A...536A..77B} Bitsch, B., Kley, W.\ 2011.\ Range of outward migration and influence of the disc's mass on the migration of giant planet cores.\ Astronomy and Astrophysics 536, A77. 

\bibitem[Borderies and Goldreich(1984)]{1984CeMec..32..127B} Borderies, N., 
Goldreich, P.\ 1984.\ A simple derivation of capture probabilities for the 
J + 1 : J and J + 2 : J orbit-orbit resonance problems.\ Celestial 
Mechanics 32, 127-136. 

\bibitem[Callegari and Yokoyama(2007)]{2007CeMDA..98....5C} Callegari, N., 
Yokoyama, T.\ 2007.\ Dynamics of two satellites in the 2/1 Mean Motion 
resonance: application to the case of Enceladus and Dione.\ Celestial 
Mechanics and Dynamical Astronomy 98, 5-30. 

\bibitem[Chambers(1999)]{1999MNRAS.304..793C} Chambers, J.~E.\ 1999.\ A 
hybrid symplectic integrator that permits close encounters between massive 
bodies.\ Monthly Notices of the Royal Astronomical Society 304, 793-799. 

\bibitem[Chirikov(1979)]{1979PhR....52..263C} Chirikov, B.~V.\ 1979.\ A 
universal instability of many-dimensional oscillator systems.\ Physics 
Reports 52, 263-379. 

\bibitem[Correia et 
al.(2009)]{2009A&A...496..521C} Correia, A.~C.~M., and 10 colleagues 2009.\ The HARPS search for southern extra-solar planets. XVI. HD 45364, a pair of planets in a 3:2 mean motion resonance.\ Astronomy and Astrophysics 496, 521-526. 

\bibitem[Cresswell and 
Nelson(2008)]{2008A&A...482..677C} Cresswell, P., Nelson, R.~P.\ 2008.\ Three-dimensional simulations of multiple protoplanets embedded in a protostellar disc.\ Astronomy and Astrophysics 482, 677-690. 

\bibitem[Crida et 
al.(2007)]{2007A&A...461.1173C} Crida, A., Morbidelli, A., Masset, F.\ 2007.\ Simulating planet migration in globally evolving disks.\ Astronomy and Astrophysics 461, 1173-1183. 

\bibitem[Delisle et 
al.(2012)]{2012A&A...546A..71D} Delisle, J.-B., Laskar, J., Correia, A.~C.~M., Bou{\'e}, G.\ 2012.\ Dissipation in planar resonant planetary systems.\ Astronomy and Astrophysics 546, A71. 

\bibitem[Duncan et al.(1998)]{1998AJ....116.2067D} Duncan, M.~J., Levison, 
H.~F., Lee, M.~H.\ 1998.\ A Multiple Time Step Symplectic Algorithm for 
Integrating Close Encounters.\ The Astronomical Journal 116, 2067-2077.

\bibitem[Fabrycky et al.(2012)]{2012arXiv1202.6328F} Fabrycky, D.~C., and 
18 colleagues 2012.\ Architecture of Kepler's Multi-transiting Systems: II. 
New investigations with twice as many candidates.\ ArXiv e-prints 
arXiv:1202.6328. 

\bibitem[Fernandez and Ip(1984)]{1984Icar...58..109F} Fernandez, J.~A., Ip, 
W.-H.\ 1984.\ Some dynamical aspects of the accretion of Uranus and Neptune 
- The exchange of orbital angular momentum with planetesimals.\ Icarus 58, 
109-120. 

\bibitem[Elskens and Escande(1991)]{1991Nonli...4..615E} Elskens, Y., 
Escande, D.~F.\ 1991.\ Slowly pulsating separatrices sweep homoclinic 
tangles where islands must be small: an extension of classical adiabatic 
theory.\ Nonlinearity 4, 615-667. 

\bibitem[Goldreich(1963)]{1963MNRAS.126..257G} Goldreich, P.\ 1963.\ On the 
eccentricity of satellite orbits in the solar system.\ Monthly Notices of 
the Royal Astronomical Society 126, 257. 

\bibitem[Goldreich and Soter(1966)]{1966Icar....5..375G} Goldreich, P., 
Soter, S.\ 1966.\ Q in the Solar System.\ Icarus 5, 375-389. 

\bibitem[Goldreich and Tremaine(1980)]{1980ApJ...241..425G} Goldreich, P., 
Tremaine, S.\ 1980.\ Disk-satellite interactions.\ The Astrophysical 
Journal 241, 425-441. 

\bibitem[Henrard(1982)]{1982amdc.proc..153H} Henrard, J.\ 1982.\ The 
adiabatic invariant - The use in celestial mechanics.\ NATO ASIC Proc.~82: 
Applications of Modern Dynamics to Celestial Mechanics and Astrodynamics 
153-171. 

\bibitem[Henrard(1982)]{1982CeMec..27....3H} Henrard, J.\ 1982.\ Capture 
into resonance - an extension of the use of adiabatic invariants.\ 
Celestial Mechanics 27, 3-22. 

\bibitem[Henrard(1983)]{1983Icar...53...55H} Henrard, J.\ 1983.\ Orbital 
evolution of the Galilean satellites - Capture into resonance.\ Icarus 53, 
55-67. 

\bibitem[Henrard and Lemaitre(1983)]{1983CeMec..30..197H} Henrard, J., 
Lamaitre, A.\ 1983.\ A second fundamental model for resonance.\ Celestial 
Mechanics 30, 197-218. 

\bibitem[Henrard et al.(1986)]{1986CeMec..38..335H} Henrard, J., Milani, 
A., Murray, C.~D., Lemaitre, A.\ 1986.\ The reducing transformation and 
apocentric librators.\ Celestial Mechanics 38, 335-344. 

\bibitem[Henrard and Lemaitre(1987)]{1987Icar...69..266H} Henrard, J., 
Lemaitre, A.\ 1987.\ A perturbative treatment of the 2/1 Jovian resonance.\ 
Icarus 69, 266-279. 

\bibitem[Henrard and Caranicolas(1990)]{1990CeMDA..47...99H} Henrard, J., 
Caranicolas, N.~D.\ 1990.\ Motion near the 3/1 resonance of the planar 
elliptic restricted three body problem.\ Celestial Mechanics and Dynamical 
Astronomy 47, 99-121. 

\bibitem[Henrard(1991)]{1991pscn.proc..193H} Henrard, J.\ 1991.\ Temporary 
capture into resonance..\ Predictability, Stability, and Chaos in N-Body 
Dynamical Systems 193-196. 

\bibitem[Hori(1966)]{1966PASJ...18..287H} Hori, G.\ 1966.\ Theory of 
General Perturbation with Unspecified Canonical Variable.\ Publications of 
the Astronomical Society of Japan 18, 287. 

\bibitem[Kirsh et al.(2009)]{2009Icar..199..197K} Kirsh, D.~R., Duncan, M., 
Brasser, R., Levison, H.~F.\ 2009.\ Simulations of planet migration driven 
by planetesimal scattering.\ Icarus 199, 197-209. 

\bibitem[Ketchum et al.(2011)]{2011ApJ...726...53K} Ketchum, J.~A., Adams, 
F.~C., Bloch, A.~M.\ 2011.\ Effects of Turbulence, Eccentricity Damping, 
and Migration Rate on the Capture of Planets into Mean Motion Resonance.\ 
The Astrophysical Journal 726, 53. 

\bibitem[Ketchum et al.(2012)]{2012arXiv1211.3078K} Ketchum, J.~A., Adams, 
F.~C., Bloch, A.~M.\ 2012.\ Mean Motion Resonances in Exoplanet Systems: An 
Investigation Into Nodding Behavior.\ ArXiv e-prints arXiv:1211.3078. 

\bibitem[Laskar and Robutel(1995)]{1995CeMDA..62..193L} Laskar, J., 
Robutel, P.\ 1995.\ Stability of the Planetary Three-Body Problem. I. 
Expansion of the Planetary Hamiltonian.\ Celestial Mechanics and Dynamical 
Astronomy 62, 193-217. 

\bibitem[Laskar(1996)]{1996CeMDA..64..115L} Laskar, J.\ 1996.\ Large Scale 
Chaos and Marginal Stability in the Solar System.\ Celestial Mechanics and 
Dynamical Astronomy 64, 115-162. 

\bibitem[Laskar(1997)]{1997A&A...317L..75L} Laskar, J.\ 1997.\ Large scale chaos and the spacing of the inner planets..\ Astronomy and Astrophysics 317, L75-L78.

\bibitem[Laskar and 
Bou{\'e}(2010)]{2010A&A...522A..60L} Laskar, J., Bou{\'e}, G.\ 2010.\ Explicit expansion of the three-body disturbing function for arbitrary eccentricities and inclinations.\ Astronomy and Astrophysics 522, A60. 

\bibitem[Lee and Peale(2002)]{2002ApJ...567..596L} Lee, M.~H., Peale, 
S.~J.\ 2002.\ Dynamics and Origin of the 2:1 Orbital Resonances of the GJ 
876 Planets.\ The Astrophysical Journal 567, 596-609. 

\bibitem[Lee et al.(2006)]{2006ApJ...641.1178L} Lee, M.~H., Butler, R.~P., 
Fischer, D.~A., Marcy, G.~W., Vogt, S.~S.\ 2006.\ On the 2:1 Orbital 
Resonance in the HD 82943 Planetary System.\ The Astrophysical Journal 641, 
1178-1187. 

\bibitem[Levison et al.(2008)]{2008Icar..196..258L} Levison, H.~F., 
Morbidelli, A., Van Laerhoven, C., Gomes, R., Tsiganis, K.\ 2008.\ Origin 
of the structure of the Kuiper belt during a dynamical instability in the 
orbits of Uranus and Neptune.\ Icarus 196, 258-273. 

\bibitem[Levison et al.(2011)]{2011AJ....142..152L} Levison, H.~F., 
Morbidelli, A., Tsiganis, K., Nesvorn{\'y}, D., Gomes, R.\ 2011.\ Late 
Orbital Instabilities in the Outer Planets Induced by Interaction with a 
Self-gravitating Planetesimal Disk.\ The Astronomical Journal 142, 152. 

\bibitem[Lin et al.(1996)]{1996Natur.380..606L} Lin, D.~N.~C., Bodenheimer, 
P., Richardson, D.~C.\ 1996.\ Orbital migration of the planetary companion 
of 51 Pegasi to its present location.\ Nature 380, 606-607. 

\bibitem[Lithwick and Wu(2012)]{2012ApJ...756L..11L} Lithwick, Y., Wu, Y.\ 
2012.\ Resonant Repulsion of Kepler Planet Pairs.\ The Astrophysical 
Journal 756, L11.

\bibitem[Malhotra(1995)]{1995AJ....110..420M} Malhotra, R.\ 1995.\ The 
Origin of Pluto's Orbit: Implications for the Solar System Beyond Neptune.\ 
The Astronomical Journal 110, 420. 

\bibitem[Mardling(2007)]{2007MNRAS.382.1768M} Mardling, R.~A.\ 2007.\ 
Long-term tidal evolution of short-period planets with companions.\ Monthly 
Notices of the Royal Astronomical Society 382, 1768-1790. 

\bibitem[Masset and Snellgrove(2001)]{2001MNRAS.320L..55M} Masset, F., 
Snellgrove, M.\ 2001.\ Reversing type II migration: resonance trapping of a 
lighter giant protoplanet.\ Monthly Notices of the Royal Astronomical 
Society 320, L55-L59. 

\bibitem[Mayor et 
al.(2004)]{2004A&A...415..391M} Mayor, M., Udry, S., Naef, D., Pepe, F., Queloz, D., Santos, N.~C., Burnet, M.\ 2004.\ The CORALIE survey for southern extra-solar planets. XII. Orbital solutions for 16 extra-solar planets discovered with CORALIE.\ Astronomy and Astrophysics 415, 391-402. 

\bibitem[Morbidelli and Moons(1993)]{1993Icar..103...99M} Morbidelli, A., 
Moons, M.\ 1993.\ Secular resonances inside mean motion commensurabilities: 
The 2/1 and 3/2 cases.\ Icarus 103, 99-108. 

\bibitem[Morbidelli(2002)]{2002mcma.book.....M} Morbidelli, A.\ 2002.\ 
Modern celestial mechanics : aspects of solar system dynamics, by Alessandro 
Morbidelli.~London: Taylor \& Francis, 2002, ISBN 0415279399 .

\bibitem[Morbidelli and Crida(2007)]{2007Icar..191..158M} Morbidelli, A., 
Crida, A.\ 2007.\ The dynamics of Jupiter and Saturn in the gaseous 
protoplanetary disk.\ Icarus 191, 158-171. 

\bibitem[Morbidelli et al.(2007)]{2007AJ....134.1790M} Morbidelli, A., 
Tsiganis, K., Crida, A., Levison, H.~F., Gomes, R.\ 2007.\ Dynamics of the 
Giant Planets of the Solar System in the Gaseous Protoplanetary Disk and 
Their Relationship to the Current Orbital Architecture.\ The Astronomical 
Journal 134, 1790-1798. 

\bibitem[Morbidelli et al.(2008)]{2008ssbn.book..275M} Morbidelli, A., 
Levison, H.~F., Gomes, R.\ 2008.\ The Dynamical Structure of the Kuiper 
Belt and Its Primordial Origin.\ The Solar System Beyond Neptune 275-292. 

\bibitem[Morbidelli et 
al.(2009)]{2009A&A...507.1041M} Morbidelli, A., Brasser, R., Tsiganis, K., Gomes, R., Levison, H.~F.\ 2009.\ Constructing the secular architecture of the solar system. I. The giant planets.\ Astronomy and Astrophysics 507, 1041-1052. 

\bibitem[Murray et al.(1998)]{1998Sci...279...69M} Murray, N., Hansen, B., 
Holman, M., Tremaine, S.\ 1998.\ Migrating Planets.\ Science 279, 69. 

\bibitem[Murray and Dermott(1999)]{1999ssd..book.....M} Murray, C.~D., 
Dermott, S.~F.\ 1999.\ Solar System Dynamics by 
Murray, C.~D., Cambridge, UK: Cambridge University 
Press 1999.

\bibitem[Murray and Holman(1997)]{1997AJ....114.1246M} Murray, N., Holman, 
M.\ 1997.\ Diffusive chaos in the outer asteroid belt..\ The Astronomical 
Journal 114, 1246-1259. 

\bibitem[Murray and Holman(1999)]{1999Sci...283.1877M} Murray, N., Holman, 
M.\ 1999.\ The Origin of Chaos in the Outer Solar System.\ Science 283, 
1877. 

\bibitem[Michtchenko et al.(2008)a]{2008MNRAS.387..747M} Michtchenko, T.~A., 
Beaug{\'e}, C., Ferraz-Mello, S.\ 2008.\ Dynamic portrait of the planetary 
2/1 mean-motion resonance - I. Systems with a more massive outer planet.\ 
Monthly Notices of the Royal Astronomical Society 387, 747-758. 

\bibitem[Neishtadt(1984)]{1984PriMM..48..197N} Neishtadt, A.~I.\ 1984.\ 
Separation of motions in systems with a rapidly rotating phase.\ 
Prikladnaia Matematika i Mekhanika 48, 197-204. 

\bibitem[Nesvorn{\'y} et al.(2002)]{2002aste.conf..379N} Nesvorn{\'y}, D., 
Ferraz-Mello, S., Holman, M., Morbidelli, A.\ 2002.\ Regular and Chaotic 
Dynamics in the Mean-Motion Resonances: Implications for the Structure and 
Evolution of the Asteroid Belt.\ Asteroids III 379-394. 

\bibitem[Nesvorn{\'y} and Morbidelli(2012)]{2012AJ....144..117N} 
Nesvorn{\'y}, D., Morbidelli, A.\ 2012.\ Statistical Study of the Early 
Solar System's Instability with Four, Five, and Six Giant Planets.\ The 
Astronomical Journal 144, 117. 

\bibitem[Paardekooper and Papaloizou(2009)]{2009MNRAS.394.2283P} 
Paardekooper, S.-J., Papaloizou, J.~C.~B.\ 2009.\ On corotation torques, 
horseshoe drag and the possibility of sustained stalled or outward 
protoplanetary migration.\ Monthly Notices of the Royal Astronomical 
Society 394, 2283-2296. 

\bibitem[Peale(1976)]{1976ARA&A..14..215P} Peale, S.~J.\ 1976.\ Orbital resonances in the solar system.\ Annual Review of Astronomy and Astrophysics 14, 215-246. 

\bibitem[Peale(1986)]{1986sate.conf..159P} Peale, S.~J.\ 1986.\ Orbital 
resonances, unusual configurations and exotic rotation statesamong 
planetary satellites..\ Satellites 159-223. 

\bibitem[Peale(1988)]{1988Icar...74..153P} Peale, S.~J.\ 1988.\ Speculative 
histories of the Uranian satellite system.\ Icarus 74, 153-171. 

\bibitem[Peale(1999)]{1999ARA&A..37..533P} Peale, S.~J.\ 1999.\ Origin and Evolution of the Natural Satellites.\ Annual Review of Astronomy and Astrophysics 37, 533-602. 

\bibitem[Poincar{\'e}(1902)]{1902BuAsI..19..289P} Poincar{\'e}, H.\ 1902.\ 
Revue des publications astronomiques. Sur les plan{\`e}tes du type 
d'h{\'e}cube.\ Bulletin Astronomique, Serie I 19, 289-310.

\bibitem[Rein et 
al.(2010)]{2010A&A...510A...4R} Rein, H., Papaloizou, J.~C.~B., Kley, W.\ 2010.\ The dynamical origin of the multi-planetary system HD 45364.\ Astronomy and Astrophysics 510, A4. 

\bibitem[Rein and 
Papaloizou(2010)]{2010A&A...524A..22R} Rein, H., Papaloizou, J.~C.~B.\ 2010.\ Stochastic orbital migration of small bodies in Saturn's rings.\ Astronomy and Astrophysics 524, A22. 

\bibitem[Rivera et al.(2005)]{2005ApJ...634..625R} Rivera, E.~J., Lissauer, 
J.~J., Butler, R.~P., Marcy, G.~W., Vogt, S.~S., Fischer, D.~A., Brown, 
T.~M., Laughlin, G., Henry, G.~W.\ 2005.\ A 7.5 M$_{\oplus}$ Planet Orbiting 
the Nearby Star, GJ 876.\ The Astrophysical Journal 634, 625-640. 

\bibitem[Sessin and Ferraz-Mello(1984)]{1984CeMec..32..307S} Sessin, W., 
Ferraz-Mello, S.\ 1984.\ Motion of two planets with periods commensurable 
in the ratio 2:1 solutions of the Hori auxiliary system.\ Celestial 
Mechanics 32, 307-332. 

\bibitem[Terquem and Papaloizou(2007)]{2007ApJ...654.1110T} Terquem, C., 
Papaloizou, J.~C.~B.\ 2007.\ Migration and the Formation of Systems of Hot 
Super-Earths and Neptunes.\ The Astrophysical Journal 654, 1110-1120. 

\bibitem[Tittemore and Wisdom(1990)]{1990Icar...85..394T} Tittemore, W.~C., 
Wisdom, J.\ 1990.\ Tidal evolution of the Uranian satellites. III - 
Evolution through the Miranda-Umbriel 3:1, Miranda-Ariel 5:3, and 
Ariel-Umbriel 2:1 mean-motion commensurabilities.\ Icarus 85, 394-443. 

\bibitem[Tsiganis et al.(2005)]{2005Natur.435..459T} Tsiganis, K., Gomes, 
R., Morbidelli, A., Levison, H.~F.\ 2005.\ Origin of the orbital 
architecture of the giant planets of the Solar System.\ Nature 435, 
459-461. 

\bibitem[Walker and Ford(1969)]{1969PhRv..188..416W} Walker, G.~H., Ford, 
J.\ 1969.\ Amplitude Instability and Ergodic Behavior for Conservative 
Nonlinear Oscillator Systems.\ Physical Review 188, 416-432. 

\bibitem[Wright et al.(2011)]{2011ApJ...730...93W} Wright, J.~T., and 10 
colleagues 2011.\ The California Planet Survey. III. A Possible 2:1 
Resonance in the Exoplanetary Triple System HD 37124.\ The Astrophysical 
Journal 730, 93.

\bibitem[Wu and Goldreich(2002)]{2002ApJ...564.1024W} Wu, Y., Goldreich, 
P.\ 2002.\ Tidal Evolution of the Planetary System around HD 83443.\ The 
Astrophysical Journal 564, 1024-1027. 

\bibitem[Wisdom(1980)]{1980AJ.....85.1122W} Wisdom, J.\ 1980.\ The 
resonance overlap criterion and the onset of stochastic behavior in the 
restricted three-body problem.\ The Astronomical Journal 85, 1122-1133. 

\bibitem[Wisdom(1983)]{1983Icar...56...51W} Wisdom, J.\ 1983.\ Chaotic 
behavior and the origin of the 3/1 Kirkwood gap.\ Icarus 56, 51-74. 

\bibitem[Wisdom(1985)]{1985Icar...63..272W} Wisdom, J.\ 1985.\ A 
perturbative treatment of motion near the 3/1 commensurability.\ Icarus 63, 
272-289. 

\bibitem[Wisdom(1986)]{1986CeMec..38..175W} Wisdom, J.\ 1986.\ Canonical 
solution of the two critical argument problem.\ Celestial Mechanics 38, 
175-180. 

\bibitem[Wisdom and Holman(1991)]{1991AJ....102.1528W} Wisdom, J., Holman, 
M.\ 1991.\ Symplectic maps for the n-body problem.\ The Astronomical 
Journal 102, 1528-1538. 

\bibitem[Yoder(1973)]{1973PhDT.........1Y} Yoder, C.~F.\ 1973.\ On the 
Establishment and Evolution of Orbit-Orbit Resonances..\ Ph.D.~Thesis . 

\bibitem[Zhang and Hamilton(2008)]{2008Icar..193..267Z} Zhang, K., 
Hamilton, D.~P.\ 2008.\ Orbital resonances in the inner neptunian system. 
II. Resonant history of Proteus, Larissa, Galatea, and Despina.\ Icarus 
193, 267-282. 

\end{thebibliography}
\end{document}